\documentstyle [12pt,amsfonts] {article}
\input epsf

\newcommand{\beq}{\begin{equation}}
\newcommand{\eeq}{\end{equation}}
\newcommand{\beqs}{\begin{eqnarray}}
\newcommand{\eeqs}{\end{eqnarray}}

\catcode`@=11
\@addtoreset{equation}{section}
\@addtoreset{equation}{subsection}
\def\theequation{\ifnum\value{section}=0 \arabic{equation}\ignorespaces
\else \ifnum\value{section}=-1 A.\arabic{equation}\ignorespaces
\else \ifnum\value{subsection}=0 \thesection.\arabic{equation}\ignorespaces
\else \thesection.\arabic{subsection}.\arabic{equation}\ignorespaces
                           \fi
                      \fi
                 \fi}
\catcode`@=12

\begin{document}

\def\thefootnote{\fnsymbol{footnote}}

\baselineskip 6.0mm

\vspace{4mm}

\begin{center}

{\Large \bf Ground State Entropy of the Potts Antiferromagnet on Strips of the
Square Lattice}

\vspace{8mm}

\setcounter{footnote}{0}
Shu-Chiuan Chang$^{(a)}$\footnote{email: shu-chiuan.chang@sunysb.edu} and
\setcounter{footnote}{6}
Robert Shrock$^{(a,b)}$\footnote{(a): permanent address;
email: robert.shrock@sunysb.edu}

\vspace{6mm}

(a) \ C. N. Yang Institute for Theoretical Physics  \\
State University of New York       \\
Stony Brook, N. Y. 11794-3840  \\

(b) \ Physics Department \\
Brookhaven National Laboratory \\
Upton, NY  11973

\vspace{10mm}

{\bf Abstract}
\end{center}

We present exact solutions for the zero-temperature partition function
(chromatic polynomial $P$) and the ground state degeneracy per site $W$ (=
exponent of the ground-state entropy) for the $q$-state Potts antiferromagnet
on strips of the square lattice of width $L_y$ vertices and arbitrarily great
length $L_x$ vertices.  The specific solutions are for (a) $L_y=4$,
$(FBC_y,PBC_x)$ (cyclic); (b) $L_y=4$, $(FBC_y,TPBC_x)$ (M\"obius); (c)
$L_y=5,6$, $(PBC_y,FBC_x)$ (cylindrical); and (d) $L_y=5$, $(FBC_y,FBC_x)$
(open), where $FBC$, $PBC$, and $TPBC$ denote free, periodic, and twisted
periodic boundary conditions, respectively.  In the $L_x \to \infty$ limit of
each strip we discuss the analytic structure of $W$ in the complex $q$ plane.
The respective $W$ functions are evaluated numerically for various values of
$q$.  Several inferences are presented for the chromatic polynomials and
analytic structure of $W$ for lattice strips with arbitrarily great $L_y$. The 
absence of a nonpathological $L_x \to \infty$ limit for real nonintegral $q$ in
the interval $0 < q < 3$ ($0 < q < 4$) for strips of the square (triangular)
lattice is discussed. 

\vspace{16mm}

\pagestyle{empty}
\newpage

\pagestyle{plain}
\pagenumbering{arabic}
\renewcommand{\thefootnote}{\arabic{footnote}}
\setcounter{footnote}{0}

\section{Introduction} 

The $q$-state Potts antiferromagnet (AF) \cite{potts,wurev} exhibits nonzero
ground state entropy, $S_0 > 0$ (without frustration) for sufficiently large
$q$ on a given lattice $\Lambda$ or, more generally, on a graph $G$.  This is
equivalent to a ground state degeneracy per site $W > 1$, since $S_0 = k_B \ln
W$.  Such nonzero ground state entropy is important as an exception to the
third law of thermodynamics \cite{cw}.  There is a close connection with 
graph theory
here, since the zero-temperature partition function of the above-mentioned
$q$-state Potts antiferromagnet on a graph $G$ satisfies
\beq
Z(G,q,T=0)_{PAF}=P(G,q)
\label{zp}
\eeq
where $P(G,q)$ is the chromatic polynomial expressing the number of ways of
coloring the vertices of the graph $G$ with $q$ colors such that no two
adjacent vertices have the same color (for reviews, see
\cite{rrev}-\cite{bbook}).  The minimum number of colors necessary for such a
coloring of $G$ is called the chromatic number, $\chi(G)$. 
Thus
\beq
W(\{G\},q) = \lim_{n \to \infty} P(G,q)^{1/n}
\label{w}
\eeq 
where $n=v(G)$ is the number of vertices of $G$ and $\{G\} = \lim_{n \to
\infty}G$.  At certain special
points $q_s$ (typically $q_s=0,1,.., \chi(G)$), one has the noncommutativity of
limits
\beq
\lim_{q \to q_s} \lim_{n \to \infty} P(G,q)^{1/n} \ne \lim_{n \to
\infty} \lim_{q \to q_s}P(G,q)^{1/n}
\label{wnoncom}
\eeq
and hence it is necessary to specify the
order of the limits in the definition of $W(\{G\},q_s)$ \cite{w}. Denoting 
$W_{qn}$ and $W_{nq}$ as the functions defined by the different order of limits
on the left and right-hand sides of (\ref{wnoncom}), we take $W \equiv W_{qn}$
here; this has the advantage of removing certain isolated discontinuities that
are present in $W_{nq}$. 

Using the expression for $P(G,q)$, one can generalize $q$ from ${\mathbb Z}_+$
to ${\mathbb C}$.  The zeros of $P(G,q)$ in the complex $q$ plane are called
chromatic zeros; a subset of these may form an accumulation set in the $n \to
\infty$ limit, denoted ${\cal B}$, which is the continuous locus of points
where $W(\{G\},q)$ is nonanalytic.~\footnote{\footnotesize{For some families of
graphs ${\cal B}$ may be null, and $W$ may also be nonanalytic at certain
discrete points.}}  The maximal region in the complex $q$ plane to which one
can analytically continue the function $W(\{G\},q)$ from physical values where
there is nonzero ground state entropy is denoted $R_1$.  The maximal value of
$q$ where ${\cal B}$ intersects the (positive) real axis is labelled
$q_c(\{G\})$.  This point is important since it separates the interval $q >
q_c(\{G\})$ on the positive real $q$ axis where the Potts model (with $q$
extended from ${\mathbb Z}_+$ to ${\mathbb R}$) exhibits nonzero ground state
entropy (which increases with $q$, asymptotically approaching $S_0 = k_B \ln q$
for large $q$, and which for a regular lattice $\Lambda$ can be calculated
approximately via large--$q$ series expansions) from the interval $0 \le q \le
q_c(\{G\})$ in which $S_0$ has a different analytic form.

  In the present work we report exact solutions for chromatic polynomials
$P(G,q)$ for strips of the square lattice with arbitrarily great length $L_x$
vertices of the following types: (a) width $L_y=4$ vertices and
$(FBC_y,PBC_x)=$ cyclic; (b) $L_y=4$ and $(FBC_y,TPBC_x)=$ M\"obius, (c)
$L_y=5$ and $(PBC_y,FBC_x)=$ cylindrical; and (d) $L_y=5$ and $(FBC_y,FBC_x)=$
open, where $FBC$, $PBC$, and $TPBC$ denote free, periodic, and twisted
periodic (i.e. periodic with reversed orientation) boundary conditions,
respectively.  For each of these, taking the infinite-length limit, we
calculate the degeneracy per site, $W(\{G\},q)$, and the continuous nonanalytic
locus ${\cal B}$.  A comparative discussion is given of these results together
with previous exact solutions for strips of smaller widths.  These strips of
regular lattices are examples of recursive families of graphs, where the latter
are constructed by successive additions of subgraph units to an initial
subgraph.  

There are several motivations for this work.  We have mentioned the basic
importance of nonzero ground state entropy in statistical mechanics \cite{cw}.
Physical examples are provided by ice \cite{lp}-\cite{liebwu} and certain other
hydrogen-bonded molecular crystals \cite{ps}.  From the point of view of
rigorous statistical mechanics, exact solutions are always valuable for the
insight that they give into the behavior of the given system under study.
Although infinite-length finite-width strips are quasi-one-dimensional systems
and hence (for finite-range spin-spin interactions) do not have
finite-temperature phase transitions, their zero-temperature critical points
are of interest.  Indeed, the presence of a zero-temperature critical point for
the Ising antiferromagnet\footnote{\footnotesize{We recall that on bipartite
graphs such as cyclic strips of the square lattice with even $L_x$, an
elementary mapping shows the Ising ferromagnet and antiferromagnet to be
equivalent; since the $L_x \to \infty$ limit can be taken with even $L_x$, this
implies that the critical behavior of the Ising ferromagnet is equivalent to
that of the Ising antiferromagnet on infinite-length limits cyclic strips of
the square lattice. By similar elementary reasoning, one can show this
equivalence for the infinite-length limit of M\"obius strips of the square
lattice.}} on infinite-length, finite-width strips of the square lattice has an
interesting connection with the behavior of the singular locus ${\cal B}$ for
the strips that we have studied with global circuits\footnote{\footnotesize{A
global circuit is a route following a lattice direction which has the topology
of the circle, $S^1$, and a length $\ell_{g.c.}$ that goes to infinity as $n
\to \infty$. For strip graphs, global circuits are equivalent to periodic or
twisted periodic boundary conditions.}}: in these cases, this singular locus
passes through the point $q=2$ in the complex $q$ plane.  Our exact solutions
for $P$ and, in the $L_x \to \infty$ limit, $W$, thus show quantitatively the
relation between critical behavior as a function of temperature (at $T=0$) in
the free energy and singularities as a function of $q$ in the per-site ground
state degeneracy $W$.  The present results also show many interesting
connections with mathematical graph theory, as is clear from the identity
(\ref{zp}), and algebraic geometry, as follows from the fact that for these
strips, ${\cal B}$ is an algebraic curve.  Besides the works already cited,
some related work on chromatic polynomials for recursive graphs includes
\cite{bds}-\cite{ss}; further discussion of background and references may be
found in \cite{a}.

A generic form for chromatic polynomials for a strip graph of type $G_s$, 
width $L_y$, and length $L_x$ is 
\beq 
P(G_s, L_y \times L_x, BC_y,BC_x,q)
= \sum_{j=1}^{N_\lambda} c_j(q)(\lambda_j(q))^{L_x}
\label{pgsum}
\eeq
where $c_j(q)$ and the $N_\lambda$ terms $\lambda_j(q)$ depend on the type of
strip graph $G_s$ including the boundary conditions but are independent of 
$L_x$.

\section{$L_y=4$ Square-Lattice Strips with $(FBC_y,(T)PBC_x)$} 

In this section we give our solutions for the chromatic polynomials of the $L_y
\times L_x$ strips of the square lattice with $(FBC_y,PBC_x)$ and
$(FBC_y,TPBC_x)$, i.e. cyclic and M\"obius, boundary conditions, respectively.
For both the cyclic and M\"obius strips, for $L_x \ge 3$ to avoid certain
degenerate cases, the square lattice strips of width $L_y$ have $n=L_xL_y$
vertices and $e=L_x(2L_y-1)$ edges.  The cyclic square strips have $\chi=2$ for
$L_x$ even and $\chi=3$ for $L_x$ odd, independent of $L_y$.  For M\"obius
square strips, $\chi=2$ for $(L_x,L_y)=(e,o)$ or $(o,e)$, and $\chi=3$ for
$(L_x,L_y)=(e,e)$ or $(o,o)$, where $e$ and $o$ denote even and odd.

We calculate the chromatic polynomials by iterated use of the
deletion-contraction theorem \cite{rtrev}, together with coloring matrix
methods \cite{b,matmeth}.  The calculation is considerably more involved than
that for the $L_y=3$ cyclic strip given in \cite{wcy}, as is indicated by the
number of $\lambda_j$ terms in eq. (\ref{pgsum}), namely, $N_\lambda=26$, as
contrasted with the value $N_\lambda=10$ for the $L_y=3$ cyclic strip.  
Elsewhere we
have given a general determination of $N_\lambda$ as a function of $L_y$
\cite{cf}.  As $L_y$ increases, the number of terms $N_\lambda$ in
(\ref{pgsum}) grows rapidly; it is 70, \ 192, \ and 534 for $L_y=5, \ 6,$ and
7.  We obtain the exact solutions of the form (\ref{pgsum})
\beq 
P(sq(4 \times m, FBC_y, PBC_x),q) = 
\sum_{j=1}^{26} c_{sq4,j} (\lambda_{sq4,j})^m
\label{pgsumly4}
\eeq
and
\beq
P(sq(4 \times m, FBC_y, TPBC_x),q) = \sum_{j=1}^{26} c_{sq4Mb,j}
(\lambda_{sq4,j})^m
\label{pgsumly4mb}
\eeq
where $L_x=m$.  
The fact that the $\lambda_j$'s for a M\"obius strip must be the same as
those for the cyclic strip of the same width and lattice type was proved in 
\cite{pm}; this also proves, {\it a fortiori}, that (i) the total number, 
$N_\lambda$, of $\lambda_j$'s, and (ii) the continuous nonanalytic locus ${\cal
B}$, including the point $q_c$, are the same for the cyclic and M\"obius strips
of a given type.  For ${\cal B}$ and $q_c$, we shall often indicate this by the
notation $(FBC_y,(T)PBC_x)$.  The explicit $\lambda_{sq,j}$'s that we calculate
are as follows.  The first six are 
\beq
\lambda_{sq4,1} = 1 
\label{lam1}
\eeq
\beq
\lambda_{sq4,2} = 3-q
\label{lam2}
\eeq
\beq
\lambda_{sq4,3} = 1-q 
\label{lam3}
\eeq
\beq
\lambda_{sq4,(4,5)}=(3 \pm \sqrt{2} \ ) - q
\label{lam45}
\eeq
and 
\beq
\lambda_{sq4,6}=(q-1)(q-3) \ . 
\label{lam6}
\eeq 
Three of the remaining $\lambda_{sq4,j}$'s, labelled $j=7,8,9$, including
the one that is dominant in region $R_1$, are identical to the three that enter
into the chromatic polynomial for the $L_y=4$ strip with $(FBC_y,FBC_x)$
boundary conditions, which was previously calculated in \cite{strip}.  This
identity was shown in \cite{bcc}.  The remaining $\lambda_{sq4,j}$'s for $10
\le j \le 26$ are roots of another cubic equation, for $j=10,11,12$; a quartic
equation for $13 \le j \le 16$; and two 5th degree equations, for $17 \le j \le
26$. Since the equations defining these $\lambda_j$'s are somewhat lengthy, we
give them in the appendix.  In Table \ref{proptable} we list various properties
of our calculation and compare them with the properties that we have found for
other related strips of the square (and triangular) lattice.  The results for
the $L_y \to \infty$ limit for the triangular lattice with $(PBC_y,FBC_x)$ are
from \cite{baxter}. Comparisons for other lattices such as honeycomb and 
kagom\'e were given in \cite{ww,strip,strip2,wcy}.

\begin{table}
\caption{\footnotesize{Properties of $P$, $W$, and ${\cal B}$ for strip graphs
$G_s$ of the square (sq) and triangular (tri) lattices.  New results in this
work are marked with an asterisk in the first column. The properties apply
for a given strip of type $G_s$ of size $L_y \times L_x$; some apply for
arbitrary $L_x$, such as $N_\lambda$, while others apply for the
infinite-length limit, such as the properties of the locus ${\cal B}$. 
For the boundary conditions in the $y$ and $x$ directions ($BC_y$,
$BC_x$), F, P, and T denote free, periodic, and orientation-reversed (twisted)
periodic, and the notation (T)P means that the results apply for either
periodic or orientation-reversed periodic. The column denoted eqs. describes
the numbers and degrees of the algebraic equations giving the
$\lambda_{G_s,j}$; for example, $\{3(1),2(2),1(3)\}$ indicates that there are 3
linear equations, 2 quadratic equations and one cubic equation.  The column
denoted BCR lists the points at which ${\cal B}$ crosses the real $q$ axis; the
largest of these is $q_c$ for the given family $G_s$. The notation ``none'' in
this column indicates that ${\cal  B}$ does not cross the real $q$ axis.  The
notation ``int;$q_1;q_c$'' refers to cases where ${\cal B}$ contains a real
interval, there is a crossing at $q_1$, and the right-hand endpoint of the
interval is $q_c$. Column labelled ``SN'' refers to whether ${\cal B}$ has
\underline{s}upport for \underline{n}egative $Re(q)$, indicated as yes (y) or
no (n).}}
\begin{center}
\begin{tabular}{|c|c|c|c|c|c|c|c|}
\hline\hline $G_s$ & $L_y$ & $BC_y$ & $BC_x$ & $N_\lambda$ & eqs. & BCR & SN
\\ \hline\hline
sq  & 1 & F    & F & 1   & \{1(1)\}      & none & n       \\ \hline
sq  & 2 & F    & F & 1   & \{1(1)\}      & none & n       \\ \hline
sq  & 3 & F    & F & 2   & \{1(2)\}      & 2    & n       \\ \hline
sq  & 4 & F    & F & 3   & \{1(3)\}      & int; \ 2.265; \ 2.30 & n \\ \hline
**sq& 5 & F    & F & 7   & \{1(7)\}      & 2.43           & n \\ \hline\hline
sq  & 1 & F &    P & 2   & \{2(1)\}    & 2, \ 0           & n \\ \hline
sq  & 2 & F & (T)P & 4   & \{4(1)\}    & 2, \ 0           & n \\ \hline   
sq  & 3 & F & (T)P & 10  & \{5(1),1(2),1(3)\} & 2.34, \ 2,\ 0 & y \\ \hline
**sq& 4 & F & (T)P & 26  &  \{4(1),1(2),2(3),1(4),2(5)\} & 2.49, \ 2, \ 0 & y 
\\ \hline\hline 
sq  & 3 & P    & F & 1   & \{1(1)\} & none & n                \\ \hline
sq  & 4 & P    & F & 2   & \{1(2)\} & int;\ 2.30;\ 2.35 & n   \\ \hline
**sq& 5 & P    & F & 2   & \{1(2)\} & 2.69  & n           \\ \hline
**sq& 6 & P    & F & 5   & \{1(5)\} & 2.61  & y           \\ \hline\hline
sq  & 3 & P    & P & 8   & \{8(1)\} & 3, \ 2, \ 0  & n    \\ \hline
sq  & 3 & P    & TP & 5   & \{5(1)\} & 3, \ 2, \ 0 & n    \\ \hline\hline
tri & 2 & F   & F & 1  & \{1(1)\} & none   & $-$          \\ \hline
tri & 3 & F   & F & 2  & \{1(2)\} & 2.57   & n            \\ \hline
tri & 4 & F   & F & 4  & \{1(4)\} & none   & n            \\ \hline
tri& 5 & F   & F & 9  & \{1(9)\} & 3     & n            \\ \hline\hline
tri & 2 & F & (T)P & 4  & \{2(1),1(2)\} & 3, \ 2, \ 0 & n  \\ \hline
tri & 3 & F & (T)P & 10 & \{3(1),2(2),1(3)\} & 3, \ 2, \ 0 & n \\ \hline
tri & 4 & F &    P & 26 & \{1(1),2(4),1(8),1(9)\} & 3.23, \ 3, \ 2, \ 0 & y \\ 
\hline\hline
tri & 3 & P   & F  & 1  & \{1(1)\}  & none  & n             \\ \hline
tri & 4 & P   & F  & 2  & \{1(2)\}  & 4, \ 3.48   & n        \\ \hline
tri & 5 & P   & F  & 2  & \{1(2)\}  & 3.28, \ 3.21 & n       \\ \hline
tri&$\infty$&P& F  & $-$ & $-$ & 4, \ 3.82, \ 0 & y  \\ \hline\hline
tri & 3 & P    & P & 11 & \{5(1),3(2)\} & 3.72, \ 2, \ 0 & n \\ \hline
tri & 3 & P   & TP & 5  & \{5(1)\}      & 3.72, \ 2, \ 0 & n \\ \hline\hline
\end{tabular}
\end{center}
\label{proptable}
\end{table}

In particular, the fact that for this width, $L_y=4$ for the cyclic strip of
the square lattice, we encounter equations of degree 5 for the $\lambda_j$'s
means that it is not possible to solve for the corresponding $\lambda_j$'s as
algebraic roots. Our experience with lattice strips of a given width $L_y$ (and
arbitrary length) and a given set of boundary conditions is that the maximal
degrees of the factors in the general equation for the $\lambda_j$'s are
non-decreasing functions of $L_y$.  Thus, assuming that this property of
non-decreasing degrees of algebraic factors in the equation for the
$\lambda_j$'s continues for higher $L_y$, our present results indicate that the
exact solutions in \cite{wcy} for the $\lambda_j$'s for the width $L_y=3$ strip
of the square lattice have completed the program of obtaining exact algebraic
expressions for these terms for this type of lattice strip.

Although no closed-form algebraic expression can be obtained for the
$\lambda_j$'s, a theorem on symmetric polynomials of roots of algebraic
equations, discussed in \cite{pm}, enables one to calculate the chromatic
polynomials exactly to arbitrary order.  The key to this is the property that
since the chromatic polynomial for a cyclic strip is a symmetric polynomial
in the various roots, it can be expressed in terms of the coefficients of the
algebraic equations that determine these $\lambda_j$'s.  However, the fact that
it is no longer possible to calculate the $\lambda_j$'s as algebraic roots when
the width of the cyclic square strip is 4 means that the determination of the
nonanalytic locus ${\cal B}$ must be done in a somewhat more cumbersome manner
than in our previous work where we had exact algebraic expressions for these
$\lambda_j$'s.

The coefficients $c_j$ that enter into the expressions for the
chromatic polynomial (\ref{pgsum}) for the cyclic and M\"obius strip of the
square lattice of width $L_y$ are certain polynomials that we denote 
$c^{(d)}$, given by \cite{cf}
\beq
c^{(d)} = \prod_{k=1}^d (q-q_{d,k})
\label{cd}
\eeq
where
\beq
q_{d,k} = 2+2\cos \Bigl ( \frac{2\pi k}{2d+1} \Bigr )
\quad {\rm for} \quad
k=1,2,...d
\label{cdzeros}
\eeq
with $0 \le d \le L_y$. We list below the specific $c^{(d)}$'s that
appear in our results for the $L_y=4$ square lattice strip:
\beq
c^{(0)}=1 \ , \quad c^{(1)}=q-1 \ , \quad c^{(2)}=q^2-3q+1 \ ,
\label{cd012}
\eeq
\beq
c^{(3)}=q^3-5q^2+6q-1 \ ,
\label{cd3}
\eeq
and
\beq
c^{(4)}=(q-1)(q^3-6q^2+9q-1) \ .
\label{cd4}
\eeq

In ascending order of degrees of $c^{(d)}$, we calculate 
\beq
c_{sq4,j}=c^{(0)} \quad {\rm for} \quad 6 \le j \le 9
\label{c6}
\eeq
\beq
c_{sq4,j}=c^{(1)} \quad {\rm for} \quad 13 \le j \le 21
\label{c1321}
\eeq
\beq
c_{sq4,j} = c^{(2)} \quad {\rm for} \quad 10 \le j \le 12 \quad {\rm and} 
\quad 22 \le j \le 26
\label{c1012}
\eeq
\beq
c_{sq4,j} = c^{(3)} \quad {\rm for} \quad 2 \le j \le 5
\label{c25}
\eeq
and
\beq
c_{sq4,1}=c^{(4)} \ . 
\label{c1}
\eeq
In \cite{bcc} it was shown that the coefficient for the $\lambda_j$ that is
leading in region $R_1$ must be 1.  

We define
\beq
C(G)=\sum_{j=1}^{N_{\lambda_G}} c_{G,j} \ .
\label{cgsum}
\eeq
where the $G$-dependence in the coefficients is indicated explicitly.  Note
that for recursive graphs like the strip graphs considered here, the $c_{G,j}$
depend on $L_y$ and the boundary conditions, but not on $L_x$. Our results
above give
\beq
C(G)=q(q-1)^3 \quad {\rm for} \quad G=sq(L_y=4,FBC_y,PBC_x) \ .
\label{csumsq}
\eeq
in accord with the generalization \cite{wcy,pm}
\beq
C(G_s(L_y \times L_x,FBC_y,PBC_x),q)=P(T_{L_y},q)=q(q-1)^{L_y-1}
\label{csumcyc}
\eeq
for $G_s$ a strip of the square (or triangular) lattice, where 
$P(T_n,q)$ is the chromatic polynomial for the tree graph $T_n$. 
This is in accord with the coloring matrix approach \cite{tk,matmeth}.

For the $L_y=4$ M\"obius strip of the square lattice, we find 
\beq
c_{sq4Mb,j}=c^{(0)} \quad {\rm for} \quad 7 \le j \le 12
\label{ccd0mb}
\eeq
\beq
c_{sq4Mb,j}=-c^{(0)} \quad {\rm for} \quad j=6 \quad {\rm and} \quad 
22 \le j \le 26
\label{cminuscd0mb}
\eeq

\beq
c_{sq4Mb,j}=c^{(1)} \quad {\rm for} \quad 17 \le j \le 21
\label{ccd1mb}
\eeq

\beq
c_{sq4Mb,j}=-c^{(1)} \quad {\rm for} \quad j=1 \quad {\rm and} \quad 
13 \le j \le 16
\label{cminuscd1mb}
\eeq
\beq
c_{sq4Mb,j}=c^{(2)} \quad {\rm for} \quad j=4,5
\label{ccd2mb}
\eeq
and
\beq
c_{sq4Mb,j}=-c^{(2)} \quad {\rm for} \quad j=2,3 \ . 
\label{cminuscd2mb}
\eeq

Hence, the sum of the coefficients is 
\beq
C(G)=0 \quad {\rm for} \quad G = sq(L_y=4,FBC_y,TPBC_x)
\label{csumvalmb}
\eeq
in accord with the general result for the M\"obius strip of the square 
(and triangular) lattice \cite{cf} 
\beq
\sum_{j=1}^{N_{\lambda_G}} 
c_{G(L_y,Mb),j} = \cases{ P(T_{\frac{L_y+1}{2}},q) & for
                                                                 odd $L_y$ \cr
                                            0 &      for even $L_y$ \cr } \ .
\label{cjsummb}
\eeq

\begin{figure}[hbtp]
\centering
\leavevmode
\epsfxsize=2.5in
\begin{center}
\leavevmode
\epsffile{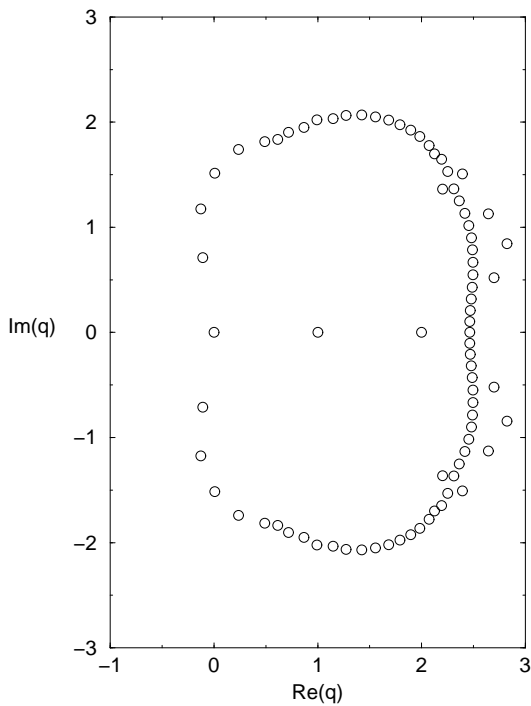}
\end{center}
\caption{\footnotesize{Chromatic zeros for the $L_y=4$,
$L_x=m=20$ (i.e., $n=80$) cyclic strip of the square lattice.}}
\label{sqpxy4}
\end{figure}

Chromatic zeros for the cyclic strip of the square lattice with $L_y=4$,
$L_x=m=20$ and hence $n=80$ are shown in Fig.  \ref{sqpxy4}; with this value of
$m$, the complex chromatic zeros lie close to the boundary ${\cal B}$
and give an approximate indication of its position. 
Note that there is a zero very close to $q=2$,
but $P(sq(L_y \times L_x,FBC_y,PBC_x),q)$ is nonzero for $q=2$ for the case
shown, where $L_x=m$ is even, as is clear from the fact that $\chi=2$ in this
case.

The maximal point at which ${\cal B}$ crosses the real axis, $q_c$, is
determined as a solution of the equation of degeneracy of leading terms
$|\lambda_{eq7-9,max}|=|\lambda_{eq22-26,max}|$, where $\lambda_{eq7-9,max}$ 
and $\lambda_{eq22-26,max}$ are the roots of eqs. (\ref{eq79}) and 
(\ref{eq2226})
with the largest magnitudes, respectively.  Since only two $\lambda_j$'s are 
degenerate in magnitude at this point, it is a regular point on the algebraic
curve ${\cal B}$ in the terminology of algebraic geometry.  This is also 
the case for the $L_y=3$ (and $L_y=1$) strip of the square lattice 
\cite{wcy,pm}, whereas, in contrast, $q_c$ is a multiple point on ${\cal B}$
for $L_y=2$.  We find 
\beq
q_c \simeq 2.4928456 \quad {\rm for} \quad sq(4 \times \infty,\ FBC_y,(T)PBC_x)
\label{qcsqcyc}
\eeq
This may be compared with the values $q_c=2$ for the $L_y \times \infty$ 
strip of the square lattice with $L_y=1,2$ and the same $(FBC_y,(T)PBC_x)$ 
boundary conditions \cite{w}, and the value 
$q_c \simeq 2.33654$ for $L_y=3$ \cite{wcy}.  We calculate that 
$W(sq,4 \times \infty,FBC_y,BC_x)=1.2697336..$ at the value $q=q_c$ in
eq. (\ref{qcsqcyc}). 

The locus ${\cal B}$ also crosses the real $q$ axis at $q=2$ and at $q=0$.  In
addition to region $R_1$ which extends outward from the envelope of ${\cal B}$
and includes the real axis for $q > q_c$ and $q < 0$, there are two other
regions that contain segments of the real axis: $R_2$, including the interval 
$2 < q < q_c$ and $R_3$, including the interval $0 < q < 2$.  In region $R_1$,
the dominant $\lambda_j$ is the root of the cubic equation (\ref{eq79}) with
the largest magnitude, which we label $\lambda_{7-9,max}$.  In region 
$R_2$, the dominant $\lambda_j$ is the root of the fifth-degree equation 
(\ref{eq2226}) with the largest magnitude, which we label 
$\lambda_{22-26, max}$.  In region $R_3$, the dominant $\lambda_j$ is the 
root of the fifth-degree equation (\ref{eq1721}) with
the largest magnitude, which we label $\lambda_{17-21,max}$.  We have
\beq
W=(\lambda_{7-9,max})^{1/4} \ , \quad {\rm for} \quad q \in R_1
\label{wr1}
\eeq

\beq
|W|=|\lambda_{22-26,max}|^{1/4} \ , \quad {\rm for} \quad q \in R_2
\label{wr2}
\eeq

\beq
|W|=|\lambda_{17-21,max}|^{1/4} \ , \quad {\rm for} \quad q \in R_3
\label{wr3}
\eeq
(In regions other than $R_1$, only the magnitude $|W|$ can be determined
unambiguously \cite{w}.) 

The locus ${\cal B}$ has support for $Re(q) < 0$ as well as $Re(q) \ge 0$. 
It separates the $q$ plane into several regions, including the three described
above and two complex-conjugate ones which we denote $R_4$ and $R_4^*$, 
centered approximately at $q \simeq 2.6 \pm 0.8i$.  In the 
regions $R_4$ and $R_4^*$, we have 
\beq
|W|=|\lambda_{13-16,max}|^{1/4} \ , \quad {\rm for} \quad q \in R_4, \ R_4^*
\label{wr4}
\eeq
Just as complex-conjugate pairs of tiny sliver regions were found for the 
cyclic $L_y=3$ square \cite{wcy} and triangular \cite{t} strips, so also 
these may be present here; we have not carried out a search for such regions
(but have ruled out the possibility of tiny regions on the real axis).

\section{$L_y=5,6$ Square-Lattice Strips with $(PBC_y,FBC_x)$}

Here we report our exact solutions for the chromatic polynomials for the width
$L_y=5,6$ strips of the square lattice of arbitrary length and with
$(PBC_y,FBC_x)$, i.e., cylindrical, boundary conditions. 
Results for the cases $L_y=3$ and $L_y=4$ were given previously in
\cite{strip2,w2d,bcc}.  We recall that $N_\lambda=1$ for $L_y=3$ and
$N_\lambda=2$ for $L_y=4$.  For $L_y=5$ and $L_y=6$ we calculate 
$N_\lambda=2$ and $N_\lambda=5$, respectively. As before, it is
convenient to present the results in terms of a generating function, denoted
$\Gamma(G_s,q,x)$.  The chromatic polynomial $P((G_s)_m,q)$ is determined as
the coefficient in a Taylor series expansion of this generating function in an
auxiliary variable $x$ about $x=0$: 
\beq 
\Gamma(G_s,q,x) =
\sum_{m=0}^{\infty}P((G_s)_m,q)x^m \ . 
\label{gamma}
\eeq
The generating function $\Gamma(G_s,q,x)$ is a rational function of the form
\beq
\Gamma(G_s,q,x) = \frac{{\cal N}(G_s,q,x)}{{\cal D}(G_s,q,x)}
\label{gammagen}
\eeq
with
\beq
{\cal N}(G_s,q,x) = \sum_{j=0}^{d_{\cal N}} A_{G_s,j}(q) x^j
\label{n}
\eeq
and
\beq
{\cal D}(G_s,q,x) = 1 + \sum_{j=1}^{d_{\cal D}} b_{G_s,j}(q) x^j
\label{d}
\eeq
where the $A_{G_s,i}$ and $b_{G_s,i}$ are polynomials in $q$, and 
$d_{\cal N} \equiv deg_x({\cal N})$, $d_{\cal D} \equiv deg_x({\cal D})$,
In factorized form 
\beq
{\cal D}(G_s,q,x) = \prod_{j=1}^{d_{\cal D}}(1-\lambda_{G_s,j}(q)x) \ . 
\label{lambdaform}
\eeq
Equivalently, the $\lambda_{G_s,j}$ are roots of the equation
\beq
\xi^{d_{\cal D}}{\cal D}(G_s,q,1/\xi) = \xi^{d_{\cal D}} + 
\sum_{j=1}^{d_{\cal D}} b_{G_s,j}\xi^{d_{\cal D}-j} \ . 
\label{xieq}
\eeq
The general formula expressing $P(G_m,q)$ in terms of these quantities is 
\cite{hs} 
\beq
P(G_m,q) = \sum_{j=1}^{d_{\cal D}} \Biggl [ \sum_{s=0}^{d_{\cal N}}
A_s \lambda_j^{d_{\cal D}-s-1} \Biggr ]
\Biggl [ \prod_{1 \le i \le d_{\cal D}; i \ne j}
\frac{1}{(\lambda_j-\lambda_i)} \Biggr ] \lambda_j^m \ . 
\label{chrompgsumlam}
\eeq

For $L_y=5$ we find 
\beq
\lambda_{sq5PF,j} = \frac{1}{2}\biggl [ T_{sq5PF} \pm
\sqrt{R_{sq5PF}} \biggr ] \ , \quad j=1,2 
\label{lamsqpy5}
\eeq
where
\beq
T_{sq5PF}=q^5-10q^4+46q^3-124q^2+198q-148
\label{tsq5pf}
\eeq
and
\beqs
& & R_{sq5PF}=q^{10}-20q^9+188q^8-1092q^7+4356q^6-12596q^5+27196q^4 \cr\cr
& & -44212q^3+52708q^2-41760q+16456 \ . 
\label{rsq5pf}
\eeqs
The coefficients $c_{sq5PF,j}$ can be computed using
eq. (\ref{chrompgsumlam}) in terms of the generating function, which is 
given in the appendix. 

In the $L_x \to \infty$ limit, the locus ${\cal B}$ includes five arcs, 
consisting of two complex-conjugate
pairs and a fifth, self-conjugate, arc.  The endpoints of
these arcs are located at the five complex-conjugate pairs of roots of
$R_{sq5PF}$.  The self-conjugate arc crosses the real axis at the real zero
of $T_{sq5PF}$, namely at 
\beq
q_c \simeq 2.691684  \quad {\rm for} \quad sq(5 \times \infty, PBC_y,FBC_x)
\label{qcsqpy5}
\eeq

\begin{figure}[hbtp]
\centering
\leavevmode
\epsfxsize=2.5in
\begin{center}
\leavevmode
\epsffile{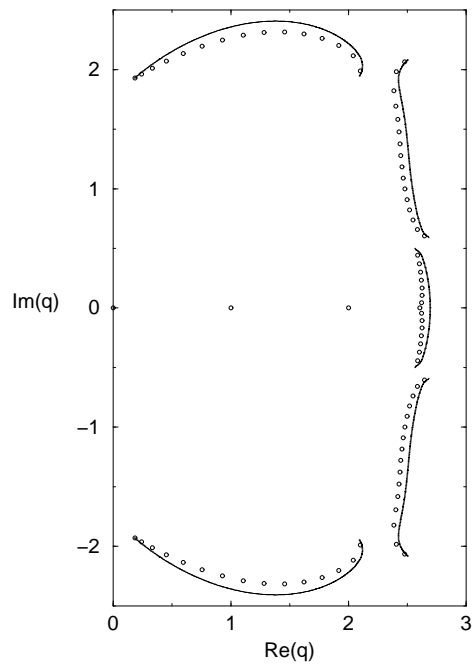}
\end{center}
\caption{\footnotesize{Locus ${\cal B}$ for the width $L_y=5$ strip (tube) of
the square lattice with $(PBC_y,FBC_x)$ boundary conditions.  Thus, the cross
sections of the tube form pentagons.  For comparison, chromatic zeros
calculated for the strip length $L_x=m+2=16$ (i.e., $n=80$ vertices) are shown.
}}
\label{sqpy5}
\end{figure}

\begin{figure}[hbtp]
\centering
\leavevmode
\epsfxsize=2.5in
\begin{center}
\leavevmode
\epsffile{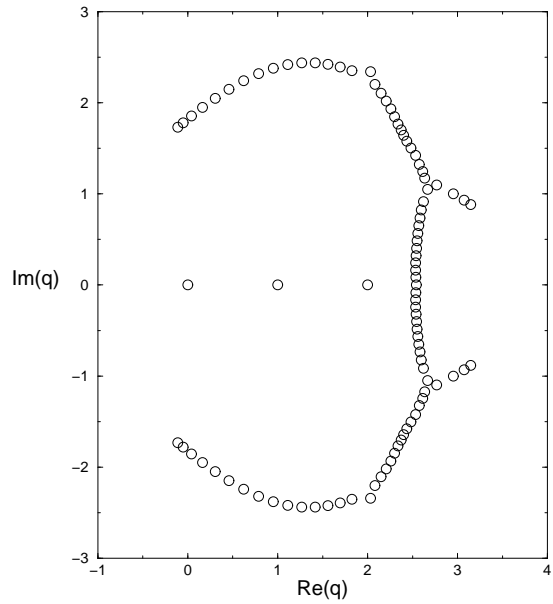}
\end{center}
\caption{\footnotesize{Locus ${\cal B}$ for the width $L_y=6$ strip (tube) of
the square lattice with $(PBC_y,FBC_x)$ boundary conditions.  Thus, the cross
sections of the tube form hexagons.  For comparison, chromatic zeros
calculated for the strip length $L_x=m+2=16$ (i.e., $n=96$ vertices) are shown.
}}
\label{sqpy6}
\end{figure}

In Fig. \ref{sqpy6} we show a plot of chromatic zeros for the $L_y=6$ strip 
of the square lattice with $(PBC_y,FBC_x)$ and length $L_x=m+2=16$ 
vertices, so that the strip has $n=96$ vertices in all. With this large a 
value of $m$, the complex
chromatic zeros lie close to the boundary ${\cal B}$ and give an approximate
indication of its position.  (We have not searched for very minute features in
${\cal B}$.) From our exact analytic results, we calculate
indication of its position.  From our exact analytic results, we calculate 
\beq
q_c \simeq 2.6089 \quad {\rm for} \quad sq(6 \times \infty,PBC_y,FBC_x)
\label{qcsqff}
\eeq 
The morphology of chromatic zeros for this long $6 \times 16$ cylindrical
strip is similar to that found for a $8 \times 8$ patch of the square lattice,
again with cylindrical boundary conditions, in \cite{baxter}.  In both cases,
the chromatic zeros have support for $Re(q) < 0$ and prongs extending to the
right; further, our exact calculation shows that in the limit $L_x \to \infty$
with $L_y=6$, the locus ${\cal B}$ has support for $Re(q) < 0$.  In
Fig. \ref{sqpy6}, one of the chromatic zeros is very close to $q=2$, but for
$q=2$ exactly, the chromatic polynomial is nonzero, equal to 2, in accord with
the fact that this strip is bipartite for any value of $L_x$.

For the $L_x \to \infty$ limit of these respective strips we have 
\beq
W(sq(5 \times \infty,PBC_y,FBC_x),q) = (\lambda_{sq5PF,j,max})^{1/5}
\label{wsqp5}
\eeq
and
\beq
W(sq(6 \times \infty,PBC_y,FBC_x),q) = (\lambda_{sq6PF,j,max})^{1/6}
\label{wsqp6}
\eeq
where $\lambda_{sq5PF,j,max}$ and $\lambda_{sq6PF,j,max}$ denote the 
solutions to the respective equations (\ref{xieq}) with maximal magnitude in
region $R_1$. 

It is of interest to use this exact result to study further the
approach of $W$ to the limit for the full infinite 2D square lattice.  This
extends our previous study in \cite{w2d}.  In Table \ref{sqpf} we list
various values of $W(sq(L_y \times \infty, PBC_y, BC_x),q)$ (which, for this
range of $q$, are independent of $BC_x$), denoted as $W(sq(L_y),P,q)$, 
together with Monte Carlo measurements of $W$ for the
full 2D square lattice, $W(sq,q)$ from \cite{ww} and the $q=3$ value 
$W(sq,3)=(4/3)^{3/2}$ from \cite{lieb}.  We also list the ratio
\beq
R_W(\Lambda(L_y),BC_y,q) = \frac{W(\Lambda(L_y),BC_y,q)}{W(\Lambda,q)}
\label{rw}
\eeq 
for the present square lattice $\Lambda=sq$.  One sees that for $L_y=5$
and moderate values of $q$, say 5 or 6, the agreement of $W(sq(L_y),q)$ for the
infinite-length, finite-width strips with the respective values $W(sq,q)$ for
the infinite square lattice is excellent; the differences are of order
$10^{-3}$ to $10^{-4}$. As noted before \cite{w2d}, for $PBC_y$ (and any
$BC_x$) this approach is not monotonic.

\begin{table}
\caption{\footnotesize{Values of $W(sq(L_y \times \infty,PBC_y,BC_x,q)$ (for
any $BC_x$), denoted $W(sq(L_y),P,q)$ for short, with $W(sq,q)=W(sq(\infty
\times \infty),q)$ for $3 \le q \le 10$.  For each value of $q$, the quantities
in the upper line are identified at the top and the quantities in the lower
line are the values of $R_W(sq(L_y),PBC_y,q)$.}}
\begin{center}
\begin{tabular}{|c|c|c|c|c|c|}
\hline\hline 
$q$ & $W(sq(3),P,q)$ & $W(sq(4),P,q)$ & $W(sq(5),P,q)$ & $W(sq(6),P,q)$ &
$W(sq,q)$ \\
\hline\hline
3 &   1.25992  & 1.58882   & 1.43097    & 1.56168    & 1.53960..  \\  
  &   0.8183   & 1.032     & 0.9294     & 1.014      & 1          \\ \hline
4 &   2.22398  & 2.37276   & 2.31865    & 2.34339    & 2.3370(7)  \\
  &   0.9516   & 1.015     & 0.9921     & 1.0027     & 1          \\ \hline
5 &   3.17480  & 3.26878   & 3.24518    & 3.25196    & 3.2510(10) \\
  &   0.9766   & 1.0055    & 0.9982     & 1.0003     & 1          \\ \hline
6 &   4.14082  & 4.21082   & 4.19790    & 4.20058    & 4.2003(12) \\
  &   0.9858   & 1.002505  & 0.9994     & 1.0001     & 1          \\ \hline
7 &   5.11723  & 5.17377   & 5.16557    & 5.16689    & 5.1669(15) \\
  &   0.9904   & 1.0013    & 0.9997     & 1.0000     & 1          \\ \hline
8 &   6.10017  & 6.14792   & 6.14221    & 6.14296    & 6.1431(20) \\
  &   0.9930   & 1.0008    & 0.9999     & 1.0000     & 1          \\ \hline
9 &   7.08734  & 7.12881   & 7.12458    & 7.12506    & 7.1254(22) \\
  &   0.9947   & 1.0005    & 0.9999     & 1.0000     & 1          \\ \hline
10 &  8.07737  & 8.11409   & 8.11083    & 8.1111     & 8.1122(25) \\
  &   0.9957   & 1.0002    & 0.9998     & 0.9999     & 1       \\ \hline\hline
\end{tabular}
\end{center}
\label{sqpf}
\end{table}

\section{$L_y=5$ Square-Lattice Strips with $(FBC_y,FBC_x)$}

We have also gone beyond the previous studies in \cite{strip,strip2} to
calculate the chromatic polynomial for the strip of the square lattice with
width $L_y=5$ and $(FBC_y,FBC_x)$, i.e., open, boundary conditions.  A related
study on wide strips is in \cite{ss}. In
\cite{strip}, a given strip $(G_s)_m$ was constructed by $m$ successive
additions of a subgraph $H$ to an endgraph $I$; here, $I=H$, so that, following
the notation of \cite{strip}, the total length of the strip graph $(G_s)_m$ is
$L_x=m+2$ vertices, or equivalently, $m+1$ edges in the longitudinal direction.
The results are conveniently expressed in terms of the coefficient functions in
the generating function, as discussed above.

For the width $L_y=5$ strip of the square lattice we find $deg_x({\cal D})=
N_\lambda = 7$.  The coefficient functions $b_{sq5FF,j}$ in eq. (\ref{d}) that
determine the $\lambda_{sq5FF,j}$'s via eq. (\ref{xieq}) are listed in the
appendix.  Because the $A_{sq5FF,j}$'s (cf. eq. (\ref{n})) are quite lengthy,
we do not give them here\footnote{\footnotesize{The $A_{sq5FF,j}$ are listed in
the copy of this paper in the cond-mat archive.}}.  In Table
\ref{proptable} this result is compared with the findings from the previous
calculations in \cite{strip,strip2} for narrower open strips of the square
lattice, and with strips of the triangular lattice \cite{wcy,t}. 
One observes that the equation (\ref{xieq}) defining the
$\lambda_j$'s increases in degree as $L_y$ increases for the open strips.  In
particular, because we now encounter an equation (\ref{xieq}) of degree higher
than 4 (specifically, degree 7), it is not possible to solve for the
$\lambda_j$'s as algebraic roots.  Furthermore, assuming that this increase in
degree of (\ref{xieq}) continues for greater widths $L_y$ of open strips, our
present results show that the previous calculations of the $\lambda_j$'s in
\cite{strip} up to $L_y=4$ have completed the program of calculating these
terms exactly as algebraic roots for open strips of the square lattice.  As
noted above, because of the theorem on symmetric polynomial functions of roots
an algebraic equations \cite{pm}, one can still calculate the chromatic
polynomial in terms of the coefficients of the algebraic equation for the
$\lambda_j$'s.

\begin{figure}[hbtp]
\centering
\leavevmode
\epsfxsize=2.5in
\begin{center}
\leavevmode
\epsffile{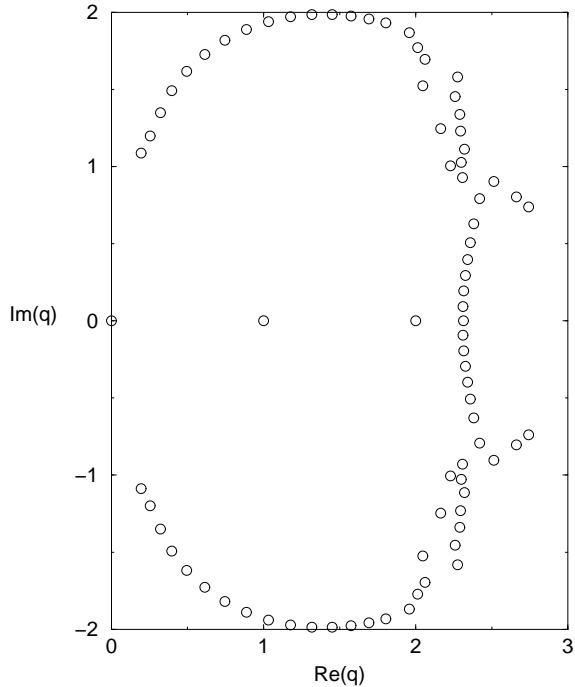}
\end{center}
\caption{\footnotesize{Chromatic zeros for the $L_y=5$ open strip of the square
lattice of length $L_x=m+2=16$ vertices (i.e. total number of vertices
$n=80$).}}
\label{sqy5}
\end{figure}

In Fig. \ref{sqy5} we show a plot of chromatic zeros for the open strip of the
square lattice with $L_y=5$ and length $L_x=m+2=16$ vertices, so that the strip
has $n=80$ vertices in all. With this large a value of $m$, the complex
chromatic zeros lie close to the boundary ${\cal B}$ and give an approximate 
indication of its position.  From an analysis of the degeneracy of leading
$\lambda_j$'s, we find that (in the $L_x \to \infty$ limit where ${\cal B}$ is
defined) 
\beq 
q_c \simeq 2.42843 \quad {\rm for} \quad sq(5
\times \infty,FBC_y,FBC_x)
\label{qctriff}
\eeq 
This is in agreement with the chromatic zeros shown in Fig. \ref{sqy5}.
Comparing Fig. \ref{sqy5} with the corresponding plots for $L_y=2$ and $L_y=3$
(Fig. 3(a,b) of \cite{strip}), we see that the arcs forming ${\cal B}$ are
elongating and that the arc endpoints nearest to the origin are approaching
more closely to the origin.  This agrees with the behavior that we had observed
earlier from narrower strips and with the conclusions that were drawn from that
behavior \cite{strip,strip2,hs,bcc}, in particular, the statement that these
results are consistent with, and provide further support for, the hypothesis
that in the limit as $L_y \to \infty$, the locus ${\cal B}$ will extend all the
way through the origin of the $q$ plane and will separate this plane into
different regions containing the real axis.

For $q > q_c$, we have, for the physical ground state degeneracy per site of
the $q$-state Potts antiferromagnet, 
\beq
W(sq(5 \times \infty,FBC_y, FBC_x), q) = (\lambda_{sq5FF,j,max})^{1/5}
\label{wsq5}
\eeq
where $\lambda_{sq5FF,j,max}$ denotes the solution of eq. (\ref{xieq}) with
the coefficients (\ref{bsq1})-(\ref{bsq7}) that has the maximal magnitude in
region $R_1$.

As with the cylindrical strips, we can use our new exact solution for the
$L_y=5$ open square strip to study the approach of $W$ to the limit for the
infinite 2D square lattice, extending \cite{w2d}.  In Table \ref{sqff} we list
various values of $W(sq(L_y \times \infty,FBC_y, BC_x),q)$ (which, for this
range of $q$, are independent of $BC_x$), denoted as $W(sq(L_y),F,q)$, together
with Monte Carlo measurements of $W$ for the full 2D square lattice, $W(sq,q)$
from \cite{ww} and the $q=3$ value $W(sq,3)=(4/3)^{3/2}$ from \cite{lieb}.  We
also list the ratio $R_W(sq(L_y),FBC_y,q)$ defined in (\ref{rw}).  In
\cite{w2d} it was proved that for $FBC_y$ the approach of $W$ to the
$L_y=\infty$ limit is monotonic.  One sees from Table \ref{sqff} that for
$L_y=5$ and moderate values of $q$, say 5 or 6, the agreement of $W(sq(L_y),q)$
for the infinite-length, finite-width strips with the respective values
$W(sq,q)$ for the infinite square lattice is very good, accurate to a few per
cent, although the approach is somewhat slower for open strips than for
cylindrical strips. This is understandable since the condition of periodic
boundary conditions in the transverse direction minimizes finite-size effects
in this direction.

\begin{table}
\caption{\footnotesize{Values of $W(sq(L_y \times \infty,FBC_y,BC_x,q)$ (for
any $BC_x$), denoted $W(sq(L_y),F,q)$ for short, with $W(sq,q)=W(sq(\infty
\times \infty),q)$ for $3 \le q \le 10$.  For each value of $q$, the quantities
in the upper line are identified at the top and the quantities in the lower
line are the values of $R_W(sq(L_y),FBC_y,q)$.}}
\begin{center}
\begin{tabular}{|c|c|c|c|c|c|c|c|} 
\hline\hline
$q$ & $W(sq(1),F,q)$ & $W(sq(2),F,q)$ & $W(sq(3),F,q)$ & $W(sq(4),F,q)$ &  
$W(sq(5),F,q)$ & $W(sq,q)$ & \\ \hline
3 &   2    & 1.73205   & 1.65846   & 1.624945  & 1.60597    & 
1.53960..  \\ 
  & 1.299  & 1.125     & 1.077     & 1.055     & 1.043   & 
1          \\ \hline
4 &   3    & 2.64575   & 2.53800   & 2.48590   & 2.45517    & 
2.3370(7)   \\
  & 1.284  & 1.132     & 1.086     & 1.064     & 1.051      & 
1           \\ \hline
5 &   4    & 3.60555   & 3.48304   & 3.42336   & 3.38805    & 
3.2510(10)  \\
  & 1.230  & 1.109     & 1.071     & 1.053     & 1.042      &
1           \\ \hline 
6 &   5    & 4.58258   & 4.45136   & 4.38717   & 4.34910    & 
4.2003(12)  \\
  & 1.190  & 1.091     & 1.060     & 1.0445    & 1.035      & 
1           \\ \hline
7 &   6    & 5.56776   & 5.43073   & 5.36348   & 5.32353    & 
5.1669(15)  \\
  & 1.161  & 1.078     & 1.051     & 1.038     & 1.030      & 
1           \\ \hline
8 &   7    & 6.55744   & 6.41623   & 6.34677   & 6.30545    & 
6.1431(20)  \\
  & 1.1395 &  1.067    & 1.0445    & 1.033     & 1.026      & 
1           \\ \hline
9 &   8    & 7.54983   & 7.40548   & 7.33434   & 7.29199    & 
7.1254(22)  \\
  & 1.123  & 1.060     & 1.039     & 1.029     & 1.023      &
1           \\ \hline
10 &   9   & 8.54400   & 8.39720   & 8.324745  & 8.28157    &
8.1122(25)  \\ 
  & 1.109  & 1.053     & 1.035     & 1.026     & 1.021      &
1           \\ \hline\hline
\end{tabular}
\end{center}
\label{sqff}
\end{table}

\section{Comparative Discussion on $P$ and ${\cal B}$}

In this section we give a general discussion of some properties of (i) the
chromatic polynomials for cyclic lattice strips with both arbitrarily great
length and arbitrarily great width, and (ii) the loci ${\cal B}$ for the
infinite-length limit of strips of the square and triangular lattice with
various boundary conditions.  This discussion incorporates the exact solutions
given in the present work and also in our previous papers. 

\begin{enumerate}

\item 

 From our exact solutions for cyclic and M\"obius strips of the square and
 triangular lattices, we draw the following inference: for these lattice
 strips, with arbitrary $L_y$ (independent of
 $L_x$), the $\lambda_{G_s,j}$ in eq. (\ref{pgsum}) with the highest-degree
 $c^{(d)}$, namely $c^{(L_y)}$ \cite{cf} (see eq. (\ref{cd})), is 
\beq
\lambda_{G_s,1}=(-1)^{L_y}
\label{lam1general}
\eeq

\item

A second inference concerns the set of terms $\lambda_{G_s,j}$ for the cyclic
strips of the square and triangular lattice with coefficients
$c_{G_s,j}=c^{(L_y-1)}$.  There are $L_y$ of these terms $\lambda_{G_s,j}$
\cite{cf}. Let $\bar\lambda_{G_x,L_y,j}=(-1)^{L_y}\lambda_{G_s,L_y,j}$ for
$G_s=sq, \ tri$.  Then the $\bar\lambda_{sq,L_y,j}$'s and hence the
$\lambda_{sq,L_y,j}$'s with coefficients $c_{sq,L_yj}=c^{(L_y-1)}$ can be
calculated as follows.  Denote the equation whose solution is
$\bar\lambda_{sq,L_y,j}$ as $f(sq,L_y,\xi)$.  Thus, $f(sq,1,\xi)=\xi+(q-1)$ and
\beq 
f(sq,2,\xi)=f(sq,1,\xi)(\xi+q-3) \ .
\label{fsq2xi}
\eeq
The $\lambda_{sq,L_y,j}$'s for higher values of $L_y$ are then given by 
\beq
f(sq,L_y,\xi) = f(sq,L_y-1,\xi)(\xi+q-3)-f(sq,L_y-2,\xi) \quad {\rm for} 
\ \ L_y \ge 3 \ . 
\label{fsqlyxi}
\eeq
We find that in the chromatic polynomial for the cyclic strip of the square
lattice, of the $\lambda_{sq,j}$'s with
coefficient $c_{sq,j}=c^{(L_y-1)}$, (i) one is $\lambda_{sq,j}=
(-1)^{L_y}(1-q)$; (ii) if $L_y$
is even, then another is $3-q$; (iii) if $L_y=0$ mod 3, two
others are $(-1)^{L_y}(2-q)$ and $(-1)^{L_y}(4-q)$; (iv) if
$L_y=0$ mod 6, then two others are $3 \pm \sqrt{3} - q$.
(This is not an exhaustive list of special factors.)  

For cyclic strips of the triangular lattice, denote the equation 
whose solution is $\bar\lambda_{tri,L_y,j}$ as $f(tri,L_y,\xi)$.  Thus, 
$f(tri,1,\xi)=\xi+(q-1)$ and 
\beq
f(tri,2,\xi) = f(tri,1,\xi)(\xi+q-3)-(\xi-1)
\label{ftri2xi}
\eeq
The $\lambda_{tri,L_y,j}$'s for higher values of $L_y$ are then given by 
\beq
f(tri,L_y,\xi) = f(tri,L_y-1,\xi)(\xi+q-3)-\xi f(tri,L_y-2,\xi) \quad {\rm for}
\ \ L_y \ge 3
\label{ftrilyxi}
\eeq
The equations defining the $\lambda_{tri,L_y,j}$'s involve progressively 
higher degrees in $\xi$.  

It was shown in \cite{pm} that the $\lambda_{G_s,j}$'s are the same for the
cyclic and M\"obius strips of a given lattice with width $L_y$.  Therefore, for
each width $L_y$, the $\lambda_{G_s,j}$'s identified above also occur in the
respective M\"obius strips of the square and triangular lattices, although they
do not, in general, have the same coefficients $c_{G_s,j}$.

These inferences are important because they show how one can reduce the problem
of calculating the $\lambda_{G_s,j}$'s for larger-width strips graphs of cyclic
and M\"obius type from those for lower widths without recourse to the usual
iterative application of the deletion-contraction or coloring matrix methods.
That is, after having used these latter methods to obtain the chromatic
polynomials for the first few values of $L_y$, the rest can be obtained purely
algebraically, without further direct analysis of the graphs involved.  Work on
constructing the recursive formulas for the other $\lambda_{G_s,j}$'s is
currently in progress. 

\item

For the infinite-length limit of a given strip graph $G_s$, the dominant
$\lambda_{G_s,j}$ in region $R_1$ is independent of the longitudinal boundary
condition, and its coefficient is $c^{(0)}=1$ \cite{bcc}.  In particular, this
$\lambda_{G_s,j}$ is the same for $(FBC_y,(T)PBC_x)$ and $(FBC_y,FBC_x)$
boundary conditions.  When this $\lambda_{G_s,j}$ is the root of an algebraic
equation of degree higher than linear, then, for the theorem on symmetric
functions of roots of algebraic equations to apply and guarantee the polynomial
nature of $P(G_s,q)$, it is necessary and sufficient that all of the other
roots of this equation enter with the same coefficient \cite{pm}.  Hence, the
analysis given in \cite{cf} for cyclic strips of type $G_s$ that determines the
number of $\lambda_{G_s,j}$'s with a specified $c_{G_s,j}=c^{(d)}$ places, for
$d=0$, an upper bound on the number $N_\lambda$ of $\lambda_{G_s,j}$'s
that occur in the strip of type $G_s$ with $(FBC_x,FBC_y)$ boundary conditions.
In particular, for $L_y$ from 1 through 8, this number $n_P(L_y,0)$ takes the
values 1,1,2,4,9,21,51,127.  For the square lattice, one finds, for $L_y$ from
1 through the current results presented here for $L_y=5$, the values
$N_\lambda=1,1,2,3,7$, as listed in Table \ref{proptable}.  For the strips of
the triangular lattice, this upper bound is realized as an equality: for $L_y$
from 1 to 5, the open strips have $N_\lambda=1,1,2,4,9$.  The reason that the
inequality is realized as an equality for the strips of the triangular lattice
is a consequence of the different behavior of the coefficients of the square
and triangular lattice strips in the M\"obius case \cite{cf}.

\item 

For all of the strips of the square lattice containing global
circuits that we have studied, the locus ${\cal B}$ encloses regions of the $q$
plane including certain intervals on the real axis and passes through $q=0$ and
$q=2$ as well as other possible points, depending on the family.  Note that the
presence of global circuits is a sufficient, but not necessary, condition for
${\cal B}$ to enclose regions, as was shown in \cite{strip2} (see Fig. 4 of
that work).  Our present results for the square lattice are in accord with, and
strengthen the evidence for, the inference (conjecture) \cite{bcc,a} that 
\beqs 
\quad {\cal B} \supset \{q=0, \ 2 \} \ \ & & {\rm for}
\ \ sq(L_y,FBC_y,(T)PBC_x) \quad \forall \ L_y \ge 1 \ \ \cr\cr 
& & {\rm and} \ \ sq(L_y,PBC_y,(T)PBC_x) \quad \forall \ L_y \ge 3 \ .
\label{bcrossq02}
\eeqs
(For the upper line of this equation, note that the $L_y=1$ graphs with 
$(FBC_y,TPBC_x)$ and $(FBC_y,PBC_x)$ boundary conditions are identical.)

\item 

The crossing of ${\cal B}$ at the point $q=2$ for the (infinite-length limit
of) strips with global circuits nicely signals the existence of a
zero-temperature critical point in the Ising antiferromagnet (equivalent to the
Ising ferromagnet on bipartite graphs). This has been discussed in \cite{a} in
the context of exact solutions for finite-temperature Potts model partition
functions on the $L_y=2$ cyclic and M\"obius strips of the square lattice.  In
contrast, this connection is not, in general, present for strips with free
longitudinal boundary conditions since ${\cal B}$ does not, in general, pass
through $q=2$. (Of the strips with $FBC_x$ that we have studied so far, such a
crossing at $q=2$ was only found for the $L_y=3$ $(FBC_y,FBC_x)$ case, as one
can see from Table \ref{proptable}.)  Furthermore, for the strips without
global circuits, there is no indication of any motion of the respective loci
${\cal B}$ toward $q=2$ as $L_y$ increases. 

\item 

Our exact solutions show that in the limit as $L_x \to \infty$, the respective
loci ${\cal B}$ for the $W$ functions for the infinite strips of the square
lattice with the fixed values of $L_y$ considered and with (i) periodic or
twisted periodic longitudinal boundary conditions and (ii) free longitudinal
boundary conditions differ; in particular, the loci ${\cal B}$ for cases with
(i) pass through $q=2$, whereas the loci for cases with (ii) do not.  This
dependence of ${\cal B}$ on the boundary conditions means that an $n \to
\infty$ limit does not exist in a manner independent of these boundary
conditions.  If one fixes $q=2$ at the outset, i.e. considers the Ising
antiferromagnet on the square-lattice strips (or if one fixes $q$ to the
trivial value $q=1$ at the outset) and then calculates $W$, there are no
pathologies; these arise when one considers nonintegral real $q$ in the range
$0 < q < 3$.  This was already discussed in the more general context of the
full temperature-dependent free energy for the Potts antiferromagnet in
\cite{a}, together with other pathologies such as a negative partition function
(lack of Gibbs measure), noted earlier in \cite{ssbounds}, and negative
specific heat.  In general, the the conclusion is that a nonpathological $n \to
\infty$ limit of the antiferromagnetic Potts model fails to exist at
sufficiently low temperature and sufficiently small real nonintegral positive
$q$ on strips of the square lattice.  Since these strips are of fixed width,
the $L_x \to \infty$ limit may be considered to be effectively
quasi-one-dimensional; in contrast, a true two-dimensional thermodynamic limit
would be $L_x \to \infty$, $L_y \to \infty$, with the ratio $L_y/L_x$ a finite
nonzero constant in this limit.  However, as is clear from the random cluster
representation of the Potts model, the problem of a negative partition function
(lack of Gibbs measure) for sufficiently small positive real nonintegral $q$ is
present for both infinite-length, finite width strips and for the above two- or
higher-dimensional infinite volume limit \cite{ssbounds,a}.  Our exact results
for infinite-length strips of various widths and our inference above that in
the $L_y \to \infty$ limit, the loci ${\cal B}$ and $W$ functions obtained with
periodic (or twisted periodic) versus free longitudinal boundary conditions
would differ is in connected with the other pathologies noted above.  From the
analysis in \cite{t}, we also conclude that a nonpathological $L_x \to \infty$
limit for the antiferromagnetic Potts model fails to exist at sufficiently low
temperature and sufficiently small positive nonintegral $q$ on strips of the
triangular lattice and a nonpathological thermodynamic limit fails to exist at
sufficiently low temperature for nonintegral $0 < q < 4$ for the full
triangular lattice.  One could infer a generalization of this for other
lattices also: a thermodynamic limit would fail to exist for the Potts
antiferromagnet at sufficiently low temperature for positive nonintegral $q$ in
the range from 0 to $q_c$ for the given 2D lattice, e.g., $q_c=3$ for the
kagom\'e lattice.  Our exact solutions are consistent with the understanding
that the point $q_c$ for the infinite 2D (or higher-dimensional) lattice is
independent of the boundary conditions used to define this infinite lattice.

\item 

For cyclic strips, we note a correlation between the coefficient $c_{G_s,j}$ of
the respective dominant $\lambda_{G_s,j}$'s in regions that include intervals
of the real axis.  Before, it was shown \cite{bcc} that the $c_{G_s,j}$ of the
dominant $\lambda_{G_s,j}$ in region $R_1$ including the real intervals $q >
q_c(\{G\})$ and $q < 0$ is $c^{(0)}=1$, where the $c^{(d)}$ were given in
eqs. (\ref{cd}), (\ref{cdzeros}).  We observe further that the
$c_{G_s,j}$ that multiplies the dominant $\lambda_{G_s,j}$ in the region
containing the intervals $0 < q < 2$ is $c^{(1)}$.  For the cyclic $L_y=3$ and
$L_y=4$ strips, there is also another region containing an interval $2 \le q
\le q_c$ on the real axis, where $q_c \simeq 2.34$ and 2.49 for $L_y=3,4$; in
this region, we find that the $c_{G_s,j}$ multiplying the dominant 
$\lambda_{G_s,j}$ is $c^{(2)}$. 

\item

Our new results on cylindrical and open strips with $L_y=5$ confirm and extend
various features that had been discussed earlier \cite{strip,strip2,hs}: for
these values of $L_y$, ${\cal B}$ forms arcs, and as $L_y$ increases, these
arcs elongate and move closer together, with the arc endpoints nearest to the
origin moving toward this point.  This is consistent with the inference that in
the $L_y \to \infty$ limit, the arcs would close to form a closed boundary that
contained $q=0$ and $q=q_c(sq)=3$.  One sees this general trend in the $L_y=6$
cylindrical strip (Fig. \ref{sqpy6}). However, in contrast with the strip
graphs containing global circuits, for which the loci ${\cal B}$ contained a
region-enclosing boundary passing through $q=0$ for any $L_y$, this feature is
evidently only approached in the limit as $L_y \to \infty$ for the strips that
do not contain global circuits.  The earlier calculations of cylindrical strips
of the triangular lattice showed an example of a strip, namely the $L_y=4$
case, where ${\cal B}$ contains arcs and an self-conjugate oval on the real
axis \cite{strip2}, but for the cylindrical strips of the square lattice that
we have investigated so far, we have not yet encountered such an oval.

\item 

For the $L_x \to \infty$ limit of all of the strips of the square lattice
containing global circuits, a $q_c$ is defined, and our results for the cyclic
and M\"obius strips with widths from $L_y=1$ through $L_y=4$ indicate that
$q_c$ is a non-decreasing function of $L_y$ in these cases.  The same behavior
was found for the strips of the triangular lattice with $L_y=2$ \cite{wcy} (and
subsequently also $L_y=3,4$ \cite{t}). 
This motivated the inference (conjecture) that $q_c$ is a
non-decreasing function of $L_y$ for strips of regular lattices with
$(FBC_y,(T)PBC_x)$ boundary conditions \cite{bcc}, and our present results
strengthen the support for this inference.  Given that, as $L_y \to \infty$,
$q_c$ reaches a limit, which is the $q_c$ for the 2D lattice of the specified
type (square, triangular, etc.), this inference leads to the following 
inequality: 
\beq 
q_c(\Lambda, L_y \times \infty,BC_y,(T)PBC_x) \le q_c(\Lambda) \ . 
\label{qcineq}
\eeq
Our exact solutions show that this inequality can be saturated.  For example, 
$q_c=3$ for the $L_y=3$ torus and Klein bottle strip of the square lattice
\cite{tk}, which is equal to the $q_c$ value for the infinite 2D square 
lattice \cite{lieb}.  
In contrast, for (the $L_x \to \infty$ limit of) strips without global
circuits, the locus ${\cal B}$ does not necessarily cross the real axis, and
hence there is not necessarily any $q_c$ defined, as was shown in
\cite{strip}.  Furthermore, in these cases, even if ${\cal B}$ does cross the
real axis, so that a $q_c$ is defined, the value of $q_c$ is not a
non-decreasing function of $L_y$.  This is shown by our calculations of 
${\cal B}$ for the $L_y=4$, $L_y=5$, and $L_y=6$ strips of the triangular 
lattice with cylindrical boundary conditions in \cite{strip2}; for
these we get $q_c=4$ for $L_y=4$ but $q_c=3.28$ for $L_y=5$ and $q_c=3.25$ for 
$L_y=6$.  Similarly, for the $L_y=5$ and $L_y=6$ cylindrical strips of the 
square lattice we get $q_c=2.69$ and $q=2.61$, respectively. 

\item 

A generalized conjecture would be to consider a slab of a $d$-dimensional 
lattice $\Lambda$ of size $L_1 \times L_2 \times ... \times L_d$, and let $d-1$
of the lengths of this slab go to infinity, holding one length, which can be
chosen without loss of generality to be $L_d$, fixed and finite, and to define 
$W$ via (\ref{w}) as 
\beq
W(\Lambda, L_d \times \infty^{d-1},BC_1,...,BC_d,q) = 
\lim_{L_j \to \infty, \ j=1,...,d-1} P(\Lambda,L_1 \times ... \times L_d,
BC_1,...,BC_d,q)^{1/n}
\label{wslab}
\eeq
For each of these $W$ functions, one would consider the corresponding
continuous singular locus ${\cal B}$ and its $q_c$, for choices of the $BC_j$
and $L_d$ where this point exists.  We display the dependence of $q_c$ on these
inputs by writing it as 
$q_c(\Lambda_d, L_d \times \infty^{d-1}, BC_1, ...,BC_d)$.  Next, we define
a $W$ function for the $d$-dimensional lattice as 
\beq
W(\Lambda_d,BC_1,...,BC_d,q) = \lim_{L_j \to \infty, \ j=1,...,d} 
P(\Lambda,L_1 \times ... \times L_d,BC_1,...,BC_d,q)^{1/n}
\label{wd}
\eeq
and a corresponding singular locus and $q_c(\Lambda_d)$.  As indicated in the
notation, one expects that this $q_c$ would be independent of the $BC_j$,
$j=1,...,d$ just as is the case for the exactly known $q_c$ values for certain
2D lattices.  Then we conjecture the inequality 
\beq
q_c(\Lambda_d, L_d \times \infty^{d-1}, BC_1, ...,BC_d) \le q_c(\Lambda_d)
\label{qcineqd}
\eeq 
Similarly, a generalization of our inference that $q_c$ is a
non-decreasing function of $L_y$ for the strips with $(FBC_y,(T)PBC_x)$ would
be the conjecture that $q_c(\Lambda_d, L_d \times \infty^{d-1}, (T)PBC_1, ...,
(T)PBC_{d-1},FBC_d)$ is a non-decreasing function of $L_d$.  Our exact
solutions for strips with $(PBC_y,FBC_x)$ boundary conditions show that if one
uses periodic rather than free boundary conditions in the direction in which
the slab is finite, then the resultant $q_c$ is not, in general, a
non-decreasing function of $L_d$.

\item 

Our exact solutions for the $L_y=4$ cyclic and M\"obius strips of the square
lattice yield a singular locus ${\cal B}$ that has support for $Re(q) < 0$.  In
comparison (see Table \ref{proptable}), this was also true for the same type of
strip with $L_y=3$, while for $L_y=1,2$, ${\cal B}$ only had support for $Re(q)
\ge 0$, and the only point on ${\cal B}$ with $Re(q)=0$ was $q=0$ itself.  This
shows that for a given type of strip, increasing $L_y$ can shift the left-most
chromatic zeros and, in the $L_x \to \infty$ limit, the left-most portion of
the locus ${\cal B}$,into the $Re(q) < 0$ half-plane.  The same type of
behavior was found for the cyclic and M\"obius strips of the triangular
lattice; for $L_y=2$ and $L_y=3$, ${\cal B}$ and chromatic zeros had support
only for $Re(q) \ge 0$, while for $L_y=4$, this support extended into the
$Re(q) < 0$ region.  In \cite{strip} it was conjectured that global circuits
were a necessary condition for lattice strips to have chromatic zeros and, in
the limit $L_x \to \infty$, a locus ${\cal B}$ with support for $Re(q) < 0$.
However, this conjecture was ruled out by our exact solutions for chromatic
polynomials, $W$, and ${\cal B}$ for homeomorphic
expansions\footnote{\footnotesize{We recall two definitions from graph theory:
(i) a homeomorphic expansion of a graph is obtained by inserting one or more
degree-2 vertices on edge(s) of the graph; (ii) the girth of a graph is the
number of edges or vertices in a minimum-distance circuit.}} of lattice strips
with $(FBC_y,FBC_x)$ boundary conditions in \cite{hs}, as also by the results
for lattice strips with $(PBC_y,FBC_x)$ in \cite{baxter}.  The homeomorphic
expansions in \cite{hs} have the effect of increasing the girth of these strip
graphs, and it was found that for a given type of open strip graph, increasing
the degree of homeomorphic expansion and hence the girth shifts the left-most
chromatic zeros and, in the limit $L_x \to \infty$, the left-most portion of
${\cal B}$, farther to the left.  This is thus a different way of getting
chromatic zeros and part of ${\cal B}$ to have support for $Re(q) < 0$ than in
the present case of cyclic strips, where this result is obtained as a
consequence of increasing the width of the strip while the girth remains
constant.  We remark that for all of these families of graphs, the magnitudes
of the chromatic zeros and points $q$ on ${\cal B}$ are bounded.  Yet another
way to get chromatic zeros and ${\cal B}$ with negative real parts involves
families with unbounded chromatic zeros and loci ${\cal B}$ \cite{wa,sokal};
indeed, in \cite{wa} we constructed families where these zeros and loci ${\cal
B}$ had arbitrarily large negative $Re(q)$

\item 

There have been a number of theorems proved concerning real chromatic zeros.
An elementary result is that no chromatic zeros can lie on the negative real
axis $q < 0$, since a chromatic polynomial has alternating coefficients.  It
has also been proved that there are no chromatic zeros in the intervals $0 < q
< 1$, and $1 < q < 32/27$ \cite{jackson}.  The bound of 32/27 in \cite{jackson}
has been shown to be sharp; i.e., for any $\epsilon > 0$, there exists a graph
with a chromatic zero at $q=32/27 + \epsilon$ \cite{thomassen}. 
Based on our studies of strips of the square (and triangular) lattices with all
of the various boundary conditions considered, we make the following
observation: for such strips, we have not found any chromatic zeros, except for
the zero at $q=1$, in the interior of the disk $|q-1|=1$.  This motivates the
conjecture that for these strips, there are no chromatic zeros with $|q-1| < 1$
except for the zero at $q=1$.  Assuming that this conjecture is valid, the
bound would be a sharp bound, since the circuit graph with $n$ vertices, $C_n$,
has chromatic zeros lying precisely on the circle $|q-1|=1$ and at $q=1$
\cite{w}. 

\end{enumerate}

\section{Values of $W$ for Low Integral Values of $q$}

In previous works \cite{w,w2d,t} and sections of the present paper we have
discussed values of $W$ for various infinite-length, finite-width lattice
strips.  For infinite-length limits of strips with global circuits, where the
region(s) of the positive real axis in the interval $0 < q < q_c$ are not
analytically connected with the region $R_1$ including $q > q_c$ (and $q < 0$),
the ground state degeneracy per site, $W$, has a qualitatively different
behavior than for integer or real $q \ge q_c$.  A comparative discussion of
this was given in \cite{w} with the results available at that time, and it is
worthwhile to use our new exact solutions to study this behavior further.  In
particular, it is of interest to inquire what the values of the $W$ functions
are for the infinite-length limits of various strips of the square lattice at
the points $q=0$, 1, and 2. Our exact analytic expressions yield the numerical
values listed in Table \ref{wsqinternal}.  The notation follows that in Table
\ref{proptable}.  As was noted in \cite{w}, in general, for regions other than
$R_1$, it is only possible to determine $|W|$ unambiguously.  Hence, for
uniformity, we list $|W|$ for all of the strips, including those with only a
region $R_1$.  For comparison, we also include values of $|W|$ at $q=0,1,2$ and
3 for infinite-length strips of the triangular lattice in Table
\ref{wtriinternal}.  In addition, for families where, in the $L_x \to \infty$
limit, there exists a $q_c$, we include the respective values of $W$ at $q_c$.
For the smallest widths, the $|W|$ values are relatively simple analytic
expressions, e.g., for the square strips, $|W|=3^{1/2}$ for $(FBC_y,BC_x)$,
$L_y=2$, $q=0$; $|W|=13^{1/3}$ for $(PBC_y,BC_x)$, $L_y=3$, $q=0$, and so
forth.  In the case of the triangular lattice, $L_y=\infty$, $(PBC_y,FBC_x)$,
the values of $|W|$ for $q=0$ and $q=4$ are from \cite{baxter}; the exact value
for $q=3$ is our analytic evaluation, and the numerical values for $q=1, \ 2$
are our numerical evaluations, of an integral representation in \cite{baxter}.
Although we list the values in the tables only to three significant figures, we
note that the $q=1$ value, $|W(tri)| \simeq 3.1716$, is different from the
$q=0$ value $|W(tri,3 \times \infty,PBC_y,BC_x)| \simeq 3.1748$.

For these values of $q$, the noncommutativity of eq. (\ref{wnoncom}) occurs
\cite{w}.  Thus, for any connected graph $G$, and in particular, the lattice
strips considered here, the chromatic polynomial $P(G,q)$ vanishes at $q=0$ and
$q=1$ and hence also the function $W_{nq}$ defined via the order of limits on
the right-hand side of eq. (\ref{wnoncom}) vanishes.  In contrast, in general,
$W_{qn}$ defined by the limits on the left-hand side of (\ref{wnoncom}) is
nonzero.  For cyclic strips of the square lattice of length $L_x$, at $q=2$,
the chromatic polynomial $P$ is equal to 2 if $L_x$ is even but 0 if $L_x$ is
odd, so that at $q=2$, no $W_{nq}$ is defined, since the limit on the
right-hand side of (\ref{wnoncom}) does not exist; however, $W_{qn}$ is
well-defined and, in general, nonzero.  Analogous comments apply for strips of
the triangular lattice: at $q=2$, the chromatic polynomial $P$ vanishes
identically, so $W_{nq}=0$, but $W_{qn}$ is, in general, nonzero.  For cyclic
strips of the triangular lattice, at $q=3$, then $P=3!$ if $L_x=0$ mod 3, and
$P=0$ if $L_x=1$ or 2 mod 3; hence, no $W_{nq}$ is defined, since the limit on
the right-hand side of (\ref{wnoncom}) does not exist, but $W_{qn}$ is
well-defined and, in general, nonzero.  As with the other results given in this
paper, the values of $W$ given in Tables \ref{wsqinternal} and
\ref{wtriinternal} follow the definition $W \equiv W_{qn}$.

\begin{table}
\caption{\footnotesize{Values of $W(sq,L_y \times \infty,BC_y,BC_x,q)$ for low
integral $q$ and for respective $q_c$, if such a point exists, where $BC_y$ and
$BC_x$ denote the transverse and longitudinal boundary conditions.}}
\begin{center}
\begin{tabular}{|c|c|c|c|c|c|c|c|}
\hline\hline 
$L_y$ & $BC_y$ & $BC_x$ & $|W_{q=0}|$ & $|W_{q=1}|$ & $|W_{q=2}|$ & $q_c$ & 
$W_{q=q_c}$ \\
\hline\hline
1  & F & F    & 1      & 0        & 1     & n      & $-$    \\ \hline
2  & F & F    & 1.73   & 1        & 1     & n      & $-$    \\ \hline
3  & F & F    & 2.06   & 1.44     & 1     & 2      & 1      \\ \hline
4  & F & F    & 2.24   & 1.68     & 1.25  & 2.30   & 1.14   \\ \hline
5  & F & F    & 2.36   & 1.82     & 1.39  & 2.43   & 1.22   \\
\hline\hline
1  & F & P    & 1      & 1       & 1      & 2      & 1     \\ \hline
2  & F & (T)P & 1.73   & 1.41    & 1      & 2      & 1     \\ \hline
3  & F & (T)P & 2.06   & 1.66    & 1.26   & 2.34   & 1.18  \\ \hline
4  & F & (T)P & 2.24   & 1.81    & 1.41   & 2.49   & 1.27  \\ 
\hline\hline
3  & P & F    & 2.35   & 1.59    & 1      & n      & $-$    \\ \hline
4  & P & F    & 2.58   & 1.89    & 1.32   & 2.35   & 1.16   \\ \hline
5  & P & F    & 2.68   & 2.05    & 1.51   & 2.69   & 1.15   \\ \hline
6  & P & F    & 2.73   & 2.14    & 1.62   & 2.61   & 1.39   \\
\hline\hline
3  & P & (T)P & 2.35   & 1.91    & 1.44   & 3      & 1.26   \\ 
\hline\hline
\end{tabular}
\end{center}
\label{wsqinternal}
\end{table}

\begin{table}
\caption{\footnotesize{Values of $W(tri,L_y \times \infty,BC_y,BC_x,q)$ for 
low integral $q$ and for respective $q_c$, if such a point exists, where 
$BC_y$ and $BC_x$ denote the transverse and longitudinal boundary
conditions.}}
\begin{center}
\begin{tabular}{|c|c|c|c|c|c|c|c|c|}
\hline\hline 
$L_y$ & $BC_y$ & $BC_x$ & $|W_{q=0}|$ & $|W_{q=1}|$ & $|W_{q=2}|$ & 
$|W_{q=3}|$ & $q_c$ & $W_{q=q_c}$ \\
\hline\hline
2  & F & F    & 2      & 1        & 0     & 1      & n     & $-$   \\ \hline
3  & F & F    & 2.49   & 1.66     & 1     & 1      & 2.57  & 0.656 \\ \hline
4  & F & F    & 2.77   & 2.02     & 1.41  & 1      & n     & $-$   \\ \hline
5  & F & F    & 2.95   & 2.25     & 1.66  & 1      & 3     & 1     \\
\hline\hline
2  & F & (T)P & 2      & 1.62    & 1      & 1      & 3     & 1     \\ \hline
3  & F & (T)P & 2.49   & 1.99    & 1.44   & 1      & 3     & 1     \\ \hline
4  & F & (T)P & 2.77   & 2.22    & 1.68   & 1.21   & 3.23  & 1.13  \\ 
\hline\hline
3  & P & F    & 3.17   & 2.22    & 1.26   & 1      & n     & $-$    \\ \hline
4  & P & F    & 3.44   & 2.60    & 1.78   & 0      & 4     & 1.19   \\ \hline
5  & P & F    & 3.56   & 2.79    & 2.05   & 1.15   & 3.28  & 0.772  \\ \hline
6  & P & F    & 3.63   & 2.90    & 2.21   & 1.41   & 3.25  & 1.11   \\ \hline
$\infty$ &P& F& 3.77   & 3.17    & 2.60   & 2      & 4     & 1.46   \\
\hline\hline
3  & P & (T)P & 3.17   & 2.62    & 2      & 1.71   & 3.72  & 1.41   \\ 
\hline\hline
\end{tabular}
\end{center}
\label{wtriinternal}
\end{table}

Some general comments follow:

\begin{enumerate}

\item 

As is evident in Tables \ref{wsqinternal} and \ref{wtriinternal}, for values of
$q$ that are positive but sufficiently small, for a given lattice, boundary
conditions, and value of $L_y$ studied, $|W|$ is a non-increasing function of
$q$.  In contrast, for sufficiently large $q$, $|W|$ increases with $q$. 
For families of graphs that involve global circuits, these two different types
of behavior occur, respectively, for $0 < q < q_c$ and $q > q_c$.  
The latter behavior is the one expected for the $q$-state Potts
antiferromagnet, since increasing $q$ increases the physical ground state
entropy. As examples, for the ($L_x \to \infty$ limit of) circuit graph, $W$ is
constant for $0 \le q \le 2$, while for the cyclic or M\"obius strip of the
square lattice $L_y=2$, it decreases as $|W|=|3-q|$ in this interval; in both
of these cases, $q_c=2$ and $W$ is real and increasing for $q > q_c$.  For the
cyclic or M\"obius $L_y=3$ and $L_y=4$ strips of the square lattice, $|W|$ has
a different analytic form in the interval $0 \le q \le 2$ and $2 \le q \le q_c$
but is everywhere decreasing for $0 < q < q_c$, for the respective values of
$q_c$.  As an example of a strip with no $q_c$, for the open line, $L_y=1$,
$|W|$ decreases from 1 to 0 as $q$ increases from 0 to 1 and increases for
larger $q$. As another example of a strip with no $q_c$, for the $L_y=3$ strip
of the square lattice with $(PBC_y,FBC_x)$ boundary conditions, $|W|$ decreases
monotonically as $q$ increases from 0 and vanishes at $q \simeq 2.453$; for
larger values of $q$, $W$ is real and positive and increases with $q$.

\item 

For the $L_x \to \infty$ limit of strips with free transverse boundary
conditions, $FBC_y$ and any longitudinal boundary conditions $BC_x$, it was
proved that for a fixed physical $q \ge q_c$, $W$ is a monotonically decreasing
function of $L_y$ \cite{w2d}.  However, as is evident from Tables
\ref{wsqinternal} and \ref{wtriinternal} for sufficiently small positive values
of $q$ (smaller than $q_c$ for strips with a $q_c$), $|W|$ is a non-decreasing
function of $L_y$. 

\item

For the strips that we have studied whose $L_x \to \infty$ limit yields a locus
${\cal B}$ with a $q_c$, $|W(q)|$ for fixed $q \in [0,q_c]$ is a non-decreasing
function of $L_y$.

\item 

It has been shown that for physical values of $q$ in the $q$-state Potts 
antiferromagnet, in the $L_x \to \infty$ limit of a strip of a given type of 
lattice $\Lambda$, $W(\Lambda, L_y \times \infty, BC_y,BC_x,q)$ is independent
of the longitudinal boundary condition $BC_x$ \cite{bcc}.  However, for small
positive values of $q$, $|W|$ does depend on both $BC_y$ and $BC_x$, as is
evident from Tables \ref{wsqinternal} and \ref{wtriinternal}.  One observes
that for the small integral values of $q$ shown in these tables, 
$|W(\Lambda, L_y \times \infty,FBC_y,FBC_x,q)|  \ \le \ 
 |W(\Lambda, L_y \times \infty,FBC_y,(T)PBC_x,q)|$ and 
$|W(\Lambda, L_y \times \infty,PBC_y,FBC_x,q)|  \ \le \ 
 |W(\Lambda, L_y \times \infty,PBC_y,(T)PBC_x,q)|$. 

\item 

It was observed in \cite{w} and proved in (section 7 of) \cite{wn} that for
integer, and, by analytic continuation, real, values of $q > max(q_c)$ for 
the square and triangular lattice strips, i.e., $q \ge 4$, $W(tri,q) <
W(sq,q)$.  Most of the values of $|W|$ shown in Tables \ref{wsqinternal} and
\ref{wtriinternal} show the opposite inequality.  Together with various other
properties noted above, this shows that $|W|$ behaves qualitatively differently
for sufficiently small positive values of $q$ than for larger values.

\end{enumerate}

\section{Conclusions}

In conclusion, we have presented exact solutions of the zero-temperature
partition function (chromatic polynomial $P$) and the ground state degeneracy
per site $W$ (= exponent of the ground-state entropy) for the $q$-state Potts
antiferromagnet on strips of the square lattice of width $L_y$ vertices and
arbitrarily great length $L_x$ vertices.  The specific solutions were for (a)
$L_y=4$, $(FBC_y,PBC_x)$ (cyclic); (b) $L_y=4$, $(FBC_y,TPBC_x)$ (M\"obius);
(c) $L_y=5,6$, $(PBC_y,FBC_x)$ (cylindrical); and (d) $L_y=5$, $(FBC_y,FBC_x)$
(open), where $FBC$, $PBC$, and $TPBC$ denote free, periodic, and twisted
periodic boundary conditions, respectively.  Some inferences were given for
certain terms $\lambda_{G_s,j}$ for cyclic and M\"obius strip graphs of the
square and triangular lattice that allow one to calculate them for arbitrarily
wide strips (of any length).  These are important because they show how one can
reduce the problem of calculating the $\lambda_{G_s,j}$'s for these strips of
arbitrarily large width from those for lower widths without recourse to the
usual iterative application of the deletion-contraction or coloring matrix
methods.  A comparative discussion was given of the continuous nonanalytic
locus ${\cal B}$ for these strips and numerical results of $W$ were given for a
range of values of $q$.  In general, our exact solutions give further
insight into the properties of the Potts antiferromagnet in the setting of
infinite-length, finite width systems.

\vspace{10mm}

Acknowledgment: The research of R. S. was supported in part by the NSF 
grant PHY-9722101 and at Brookhaven by the DOE contract
DE-AC02-98CH10886.\footnote{\footnotesize{Accordingly, the U.S. government
retains a non-exclusive royalty-free license to publish or reproduce the
published form of this contribution or to allow others to do so for
U.S. government purposes.}} 

\section{Appendix}

\subsection{Terms $\lambda_j$ for the Cyclic $4 \times m$  Strip of 
the Square Lattice}

In this appendix we give the equations for the terms $\lambda_j$ for $7 \le j
\le 26$.  The $\lambda_{sq4,j}$, $j=7,8,9$, are roots of the cubic equation
\beqs
& & \xi^3 + (-q^4+7q^3-23q^2+41q-33)\xi^2  \cr\cr 
& & + (2q^6-23q^5+116q^4-329q^3+553q^2-517q+207)\xi  \cr\cr
& & + (-q^8+16q^7-112q^6+449q^5-1130q^4+1829q^3-1858q^2+1084q-279) = 0 \ . 
\cr\cr
& & 
\label{eq79}
\eeqs
The $\lambda_j$ for $j=10,11,12$ are the roots of another cubic equation
\beqs
& & \xi^3 - 2(q-2)(q-3)\xi^2 + (q^4-11q^3+42q^2-68q+38)\xi \cr\cr
& & - (q-1)(q-3)(-q^3+7q^2-15q+11)=0 \ . 
\label{eq1012}
\eeqs
The $\lambda_j$ for $13 \le j \le 16$ are the roots of the quartic equation
\beqs
& & \xi^4 + (2q^3-12q^2+28q-23)\xi^3  \cr\cr
& & + (q^6-13q^5+73q^4-224q^3+396q^2-381q+152)\xi^2  \cr\cr
& & - (q-1)(q-3)(q^6-12q^5+62q^4-179q^3+304q^2-288q+119)\xi \cr\cr
& & -q^9+17q^8-127q^7+549q^6-1518q^5+2790q^4-3411q^3+2673q^2-1215q+243=0 \ . 
\cr\cr
& & 
\label{eq1316}
\eeqs
Finally, there are two sets of roots of two degree-5 equations.  The set 
$\lambda_j$ for $17 \le j \le 21$ are the roots of the equation
\beqs
& & \xi^5 + (q-3)(2q^2-9q+14)\xi^4 \cr\cr
& & + (q^6-17q^5+119q^4-446q^3+947q^2-1080q+515)\xi^3 \cr\cr
& & - (q-3)(2q^7-32q^6+224q^5-883q^4+2106q^3-3028q^2+2428q-835)\xi^2 \cr\cr
& & + (q-2)(q^9-22q^8+213q^7-1193q^6+4267q^5-10120q^4+15922q^3-16013q^2+9329q
-2394)\xi \cr\cr
& & 
+ (q-1)(q^8-17q^7+125q^6-520q^5+1342q^4-2206q^3+2261q^2-1325q+341)(q-3)^2=0 
\ . \cr\cr
& & 
\label{eq1721}
\eeqs
The set $\lambda_j$ for $22 \le j \le 26$ are the roots of the equation
\beqs
& & \xi^5 + (-4q^2+19q-26)\xi^4 + (6q^4-58q^3+214q^2-354q+219)\xi^3 \cr\cr
& & + (-4q^6+60q^5-370q^4+1198q^3-2144q^2+2013q-773)\xi^2 \cr\cr
& & + (q-2)(q^7-20q^6+162q^5-693q^4+1697q^3-2391q^2+1805q-565)\xi \cr\cr
& & + (q-1)(q-3)(q^7-16q^6+106q^5-378q^4+788q^3-967q^2+653q-189)=0 \ . 
\cr\cr
& & 
\label{eq2226}
\eeqs

\subsection{Generating Functions for the $L_y=5,6$ Strips of the Square 
Lattice with $(PBC_y,FBC_x)$}

For the $L_y=5$ strip we calculate a generating function of the form 
(\ref{gammagen}) with $d_{\cal D}=2$, $d_{\cal N}=1$ and, in the notation of 
eqs. (\ref{d}) and (\ref{n}), we find 
\beq
b_{sq5PF,1}=-q^5+10q^4-46q^3+124q^2-198q+148
\label{bsq5pf1}
\eeq

\beq
b_{sq5PF,2}=q^8-19q^7+159q^6-767q^5+2339q^4-4627q^3+5800q^2-4212q+1362
\label{bsq5pf2}
\eeq

\beq
A_{sq5PF,0}=q(q-1)(q-2)(q^7-12q^6+67q^5-225q^4+494q^3-719q^2+650q-282)
\label{asq5pf0}
\eeq

\beqs
& & A_{sq5PF,1}=-q(q-1)(q-2)(q^2-2q+2)(q^8-19q^7+159q^6-767q^5 \cr\cr
& & +2339q^4-4627q^3+5800q^2-4212q+1362)
\label{asq5pf1}
\eeqs

For the $L_y=6$ strip we calculate a generating function of the form 
(\ref{gammagen}) with $d_{\cal D}=5$, $d_{\cal N}=4$, with 
\beq
b_{sq6PF,1}=-q^6+12q^5-68q^4+234q^3-524q^2+727q-483
\label{bsq6pf1}
\eeq

\beqs
& & b_{sq6PF,2}=2q^{10}-44q^9+456q^8-2917q^7+12710q^6-39322q^5+87323q^4
-137193q^3 \cr\cr
& & +145624q^2-94100q+28114
\label{bsq6pf2}
\eeqs

\beqs
& & b_{sq6PF,3}=-q^{14}+33q^{13}-509q^{12}+4872q^{11}-32374q^{10}+158152q^9
-586234q^8 \cr\cr
& & +1676100q^7-3715937q^6+6358772q^5-8268225q^4+7921161q^3 \cr\cr
& & -5284418q^2+2197026q-429510
\label{bsq6pf3}
\eeqs

\beqs
& & b_{sq6PF,4}=-q^{17}+38q^{16}-681q^{15}+7649q^{14}-60357q^{13}+355400q^{12}
-1618550q^{11} \cr\cr
& & +5828269q^{10}-16812727q^9+39098146q^8-73327191q^7+110295876q^6
-131415610q^5 \cr\cr 
& & +121386275q^4-83893487q^3+40850378q^2-12502528q+1809361
\label{bsq6pf4}
\eeqs

\beqs
& & b_{sq6PF,5}=q^{19}-41q^{18}+794q^{17}-9658q^{16}+82760q^{15}-531052q^{14}
\cr\cr
& & +2647330q^{13}-10495556q^{12}+33592560q^{11}-87588439q^{10}+186851845q^9 
-326185418q^8 \cr\cr
& & +464098186q^7-533530852q^6+488389118q^5-347889815q^4+185960167q^3
-70211630q^2 \cr\cr
& & +16703951q-1884267
\label{bsq6pf5}
\eeqs

\beqs
& & A_{sq6PF,0}=q(q-1)(q^{10}-17q^9+136q^8-674q^7+2296q^6-5640q^5+10183q^4
-13457q^3 \cr\cr
& & +12563q^2-7517q+2183)
\label{asq6pf0}
\eeqs

\beqs
& & A_{sq6PF,1}=-q(q-1)(2q^{14}-54q^{13}+695q^{12}-5631q^{11}+31999q^{10}
-134668q^9 \cr\cr
& &  +432404q^8-1075802q^7+2085064q^6-3137110q^5+3615627q^4-3106751q^3
+1890461q^2 \cr\cr
& & -733250q+137516)
\label{asq6pf1}
\eeqs

\beqs
& & A_{sq6PF,2}=q(q-1)(q^{18}-38q^{17}+684q^{16}-7756q^{15}+62128q^{14}
-373554q^{13} \cr\cr
& & +1748131q^{12}-6513823q^{11}+19602672q^{10}-48032023q^9+96128905q^8
-156920332q^7 \cr\cr
& & +207640116q^6-220043849q^5+183010634q^4-115543495q^3+52290297q^2 \cr\cr
& & -15182726q+2135038)
\label{asq6pf2}
\eeqs

\beqs
& & A_{sq6PF,3}=q(q-1)(q^{21}-43q^{20}+881q^{19}-11444q^{18}+105796q^{17}
-740641q^{16} \cr\cr
& & +4078480q^{15}-18111664q^{14}+65961019q^{13}-199240735q^{12}
+502713558q^{11} \cr\cr
& & -1063474616q^{10}+1887470282q^9-2803761470q^8+3465937164q^7-3530769703q^6
+2919336052q^5 \cr\cr
& & -1914246633q^4+960052617q^3-346744827q^2+80479446q-9033772)
\label{asq6pf3}
\eeqs

\beqs
& & A_{sq6PF,4}=-q(q-1)(q^4-5q^3+10q^2-10q+5)(q^{19}-41q^{18}+794q^{17}
-9658q^{16} \cr\cr
& & +82760q^{15}-531052q^{14}+2647330q^{13}-10495556q^{12}+33592560q^{11}
-87588439q^{10} \cr\cr
& & +186851845q^9-326185418q^8+464098186q^7-533530852q^6+488389118q^5 \cr\cr
& & -347889815q^4+185960167q^3-70211630q^2+16703951q-1884267)
\label{asq6pf4}
\eeqs

\subsection{Generating Function for the $L_y=5$ Open Strip of the Square 
Lattice}

For this strip we calculate a generating function of the form (\ref{gammagen})
with $d_{\cal D}=7$ and $d_{\cal N}=6$. In the notation of eq. (\ref{d}) we
find 
\beq 
b_{sq(5),1}=-q^5+9q^4-40q^3+107q^2-167q+118
\label{bsq1}
\eeq

\beqs
& &
b_{sq5FF,2}=4q^8-63q^7+458q^6-2011q^5+5840q^4-11477q^3+14844q^2-11466q+4003
\cr\cr 
& & 
\label{bsq2}
\eeqs

\beqs
& & b_{sq5FF,3}= -6q^{11}+136q^{10}-1432q^9+9250q^8-40749q^7+128594q^6 \cr\cr
& & -296624q^5+499762q^4-601803q^3+492117q^2-245164q+56113
\label{bsq3}
\eeqs

\beqs 
& & b_{sq5FF,4}=4q^{14}-120q^{13}+1685q^{12}-14681q^{11}+88695q^{10}
-393187q^9+1319323q^8 \cr\cr
& & 
-3404712q^7+6790667q^6-10414582q^5+12084263q^4-10278730q^3 \cr\cr 
& & +6051725q^2-2204111q+373840 
\label{bsq4}
\eeqs

\beqs & & b_{sq5FF,5}=-(q-1)(q^{16}-38q^{15}+674q^{14}-7419q^{13}+56807q^{12}
-321258q^{11} \cr\cr
& &
+1389731q^{10}-4696189q^9+12540817q^8-26576855q^7+44582788q^6-58613690q^5
\cr\cr 
& & +59234653q^4-44497390q^3+23436736q^2-7733009q+1204091)
\label{bsq5}
\eeqs

\beqs
& & b_{sq5FF,6}=
-(q-1)^2(q^{17}-38q^{16}+682q^{15}-7680q^{14}+60795q^{13}-359135q^{12}+
1639962q^{11} \cr\cr
& & 
-5915021q^{10}+17065698q^9-39623309q^8+74056302q^7-110813572q^6 \cr\cr
& & +131155616q^5-120231650q^4+82455281q^3-39872376q^2 \cr\cr
& & +12141916q-1753922)
\label{bsq6}
\eeqs

\beqs
& & b_{sq5FF,7}=(q-1)^3(q-2)^2(q^{15}-34q^{14}+538q^{13}-5259q^{12}+35541q^{11}
-176036q^{10} \cr\cr
& & +660682q^9-1914798q^8+4324155q^7-7615130q^6+10381339q^5-10768339q^4 \cr\cr
& & +8235159q^3-4388527q^2+1459163q-228580) \ . 
\label{bsq7}
\eeqs 
Since the $A_{sq5FF,j}$ are rather lengthy, they are given in the copy of
this paper in the cond-mat archive. With the definition 
$A_{sq5FF,j}=q(q-1)\bar A_{sq5FF,j}$, we have
\beq
\bar A_{sq5FF,0}=(D_4)^4
\label{a0}
\eeq
where $D_4=q^2-3q+3$;

\beqs
& & \bar A_{sq5FF,1}=-4q^{11}+75q^{10}-653q^9+3478q^8-12572q^7+32346q^6
\cr\cr
& & -60381q^5+81687q^4-78370q^3+50664q^2-19788q+3517
\label{asq1}
\eeqs

\beqs
& & \bar A_{sq5FF,2}=6q^{14}-154q^{13}+1854q^{12}-13864q^{11}+71883q^{10}
\cr\cr
& & -273164q^9+784036q^8-1725384q^7+2923023q^6-3789945q^5+3697547q^4 \cr\cr
& & -2627998q^3+1283656q^2-384667q+53170
\label{asq2}
\eeqs

\beqs
& &
\bar A_{sq5FF,3}=-4q^{17}+132q^{16}-2056q^{15}+20067q^{14}-137414q^{13}
\cr\cr
& & +700413q^{12}-2750993q^{11}+8502143q^{10}-20926266q^9+41238149q^8 \cr\cr
& & -65036748q^7+81574624q^6-80321984q^5+60731068q^4-34014621q^3 \cr\cr
& & +13280602q^2-3222200q+365089
\label{asq3}
\eeqs

\beqs
& &
\bar A_{sq5FF,4}=(q-1)^2(q^{18}-40q^{17}+751q^{16}-8804q^{15}+72287q^{14}
-441816q^{13} \cr\cr
& & +2084720q^{12}-7769759q^{11}+23199424q^{10}-55934061q^9+109187879q^8 \cr\cr
& & -172199452q^7+217795442q^6-217905554q^5+168626444q^4-97343549q^3 \cr\cr
& & +39444397q^2-10000178q+1191823)
\label{asq4}
\eeqs

\beqs
& &
\bar A_{sq5FF,5}=(q-1)^3(q^{19}-40q^{18}+759q^{17}-9082q^{16}+76836q^{15}
-488373q^{14} \cr\cr
& & +2418556q^{13}-9549855q^{12}+30509903q^{11}-79556288q^{10}
+169998158q^9 \cr\cr
& & -297636195q^8+425129474q^7-490934225q^6+451509111q^5-323033449q^4 \cr\cr
& & +173271795q^3-65533774q^2+15574281q-1747588)
\label{asq5}
\eeqs

\beqs
& &
\bar A_{sq5FF,6}=-(q-1)^6(q-2)^2(q^{15}-34q^{14}+538q^{13}-5259q^{12}
+35541q^{11} \cr\cr
& & -176036q^{10}+660682q^9-1914798q^8+4324155q^7-7615130q^6+10381339q^5 \cr\cr
& & -10768339q^4+8235159q^3-4388527q^2+1459163q-228580)
\label{asq6}
\eeqs

\vfill
\eject
\end{document}